\documentclass[twocolumn]{aastex63}

\usepackage{mathtools} 
\usepackage{booktabs}

\begin{document}

\newcommand*\phantomrel[1]{\mathrel{\phantom{#1}}}

\newcommand{\rev}[1]{{\color{black} #1}}

\newcommand{\pc}{\,{\rm pc}}
\newcommand{\kpc}{\,{\rm kpc}}
\newcommand{\second}{\,{\rm s}}
\newcommand{\yr}{\,{\rm yr}}
\newcommand{\Myr}{\,{\rm Myr}}
\newcommand{\cm}{\,{\rm cm}}
\newcommand{\pcc}{\,{\rm cm}^{-3}}
\newcommand{\kms}{\,{\rm km}\,{\rm s}^{-1}}
\newcommand{\Msun}{\,M_{\odot}}
\newcommand{\Lsun}{\,L_{\odot}}
\newcommand{\Kel}{\,{\rm K}}
\newcommand{\kB}{{\,k_{\rm B}}}
\newcommand{\eV}{{\,{\rm eV}}}

\newcommand{\HII}{\ion{H}{2}\ignorespaces}
\newcommand{\Halpha}{${\rm H}\alpha$}

\newcommand{\Qi}{Q_{\rm i}}
\newcommand{\nH}{n_{\rm H}}
\newcommand{\nHI}{n_{\rm H^0}}
\newcommand{\nHII}{n_{\rm H^+}}
\newcommand{\nelec}{n_{\rm e}}
\newcommand{\xn}{x_{\rm n}}
\newcommand{\alphaB}{\alpha_{\rm B}}
\newcommand{\nesq}{n_{\rm e}^2}
\newcommand{\thh}{$^{\rm th}$}

\newcommand{\Thetaw}{\Theta_{\rm w}}
\newcommand{\mean}[2]{{\langle{#1}\rangle_{#2}}}
\newcommand{\parallelsum}{\mathbin{\|}}
\newcommand{\hsep}{200 pc}

\title{Diffuse Ionized Gas in Simulations of Multiphase, Star-Forming Galactic
  Disks}

\shorttitle{DIG in Star-forming Galactic Disks} %

\shortauthors{Kado-Fong et al.} %

\author[0000-0002-0332-177X]{Erin Kado-Fong} %

\author[0000-0001-6228-8634]{Jeong-Gyu Kim} %

\author[0000-0002-0509-9113]{Eve C.~Ostriker} %

\author[0000-0003-2896-3725]{Chang-Goo Kim} %

\affiliation{Department of Astrophysical Sciences, Princeton University,
  Princeton, NJ 08544, USA}

\email{kadofong@princeton.edu, kimjg@astro.princeton.edu,
  eco@astro.princeton.edu, cgkim@astro.princeton.edu}

\date{\today}

\begin{abstract}
  It has been hypothesized that photons from young, massive star clusters are
  responsible for maintaining the ionization of diffuse warm ionized gas seen in
  both the Milky Way and other disk galaxies. For a theoretical investigation of
  the warm ionized medium (WIM), it is crucial to solve radiation transfer
  equations where the ISM and clusters are modeled self-consistently. To this
  end, we employ a Solar neighborhood model of TIGRESS, a magnetohydrodynamic
  simulation of the multiphase, star-forming ISM, and post-process the
  simulation with an adaptive ray tracing method to transfer UV radiation from
  star clusters. We find that the WIM volume filling factor is highly variable,
  and sensitive to the rate of ionizing photon production and ISM structure. The
  mean WIM volume filling factor rises to $\sim 0.15$ at $|z|\sim1 \kpc$.
  Approximately half of ionizing photons are absorbed by gas and half by dust;
  the cumulative ionizing photon escape fraction is 1.1\%. Our time-averaged
  synthetic \Halpha\ line profile matches WHAM observations on the redshifted
  (outflowing) side, but has insufficient intensity on the blueshifted side. Our
  simulation matches the Dickey-Lockman neutral density profile well, but only a
  small fraction of snapshots have high-altitude WIM density consistent with
  Reynolds Layer estimates. We compute a clumping correction factor
  $\mathcal{C}_{n_e} \equiv \langle n_e\rangle/\langle n_e^2\rangle^{1/2} \sim
  0.2$ that is remarkably constant with distance from the midplane and time;
  this can be used to improve estimates of ionized gas mass and mean electron
  density from observed \Halpha\ surface brightness profiles in edge-on
  galaxies.
\end{abstract}

\section{Introduction}\label{s:intro}

The presence of a diffuse layer of ionized gas reaching over $1 \kpc$ above the
Galactic plane has been known for decades \citep{hoyle1963, reynolds1973}, and
the physical properties of this diffuse warm ionized medium (WIM) have been
characterized based on a variety of observational diagnostics
\citep[e.g.,][]{reynolds1989, reynolds1991a, madsen2006, gaensler2008,
  hill2008}. Wide field \Halpha\ surveys have in particular expanded the view of
the WIM to the full sky \citep{haffner2003,haffner2010}. Beyond the Milky Way,
analogous distributions of diffuse ionized gas (DIG) have been seen in many
nearby disk galaxies
\citep[e.g.,][]{rand1990,dettmar1990,zurita2000,jones2017,jo2018,levy2019}.
\citet{haffner2009} reviews both Galactic and extragalactic observations, and
adopts the convention of using ``WIM'' for the Milky Way and ``DIG'' for other galaxies
to indicate a diffuse component of warm ionized gas. In this work, we use ``WIM'' or ``warm ionized gas'' 
to refer to any photoionized gas \textit{in simulations}, and 
apply the term ``DIG'' to the diffuse portion at high 
altitude regions. When referring to observations, we also use the term
``WIM'' for generic ionized gas if the context is clear.

Observations of optical line emission ratios suggest that the temperature of the WIM
ranges from $6000 \Kel$ to $10000 \Kel$, slightly higher than that of
classical \HII\ regions \citep[e.g.,][]{madsen2006}. The dispersion measure
(${\rm DM} = \int \nelec d\ell$) of pulsars with known distances shows that the
line-of-sight averaged free electron density is $\sim 0.01$--$0.1 \pcc$, with
a vertical exponential scale height of $h(\nelec) \sim 1 \kpc$ (e.g.,
\citealt{reynolds1991b,nordgren1992,gomez2001,gaensler2008,deller2019}; see also
\citealt{savage1990,peterson2002,savage2009}, which estimated the scale height
via other means). Studies combining measurements of the emission measure
(${\rm EM} = \int \nesq d\ell$; typically derived from \Halpha\ surface brightness) and the DM
suggest that the volume filling fraction of WIM is $\lesssim 0.1$ at the midplane
and $\sim 0.2$--$0.4$ at $|z| \sim 1\kpc$
\citep[e.g.,][]{reynolds1991a,berkhuijsen2006,gaensler2008}. The Milky Way EM scale height 
(obtained by fitting an exponential to the \Halpha\ intensity as a function
of height from the midplane) is smaller ($\sim 250$--$550\pc$;
\citealt{haffner1999,hill2014,krishnarao2017}), although an anomalously high value ($>1\kpc$)
has been found along the far Carina arm \citep{krishnarao2017}. The \Halpha\
scale heights of external (edge-on) galaxies range from a few hundred pc to over
$2 \kpc$ \citep[e.g.,][]{jo2018,boettcher2019,levy2019}. In both the Milky Way
and other nearby galaxies, the distribution of diffuse \Halpha\ surface
brightness (projected on the disk plane) is well characterized by a lognormal
\citep{hill2008, seon2009,berkhuijsen2015}.

The mechanism for maintaining warm ionized gas far from the midplane has been
debated since its discovery. Proposed mechanisms for the origin of extraplanar
warm gas include the cooling of hot galactic fountain gas \citep{shapiro1976,
  bregman1980}, gas accretion from the intergalactic medium \citep{binney2005},
and entrainment of warm ISM clouds by hot winds. Based on analysis of flows in
numerical simulations with clustered supernovae, \citet{kor2017} and
\citet{kim2018} showed that significant amounts of warm gas are accelerated by
superbubble expansion, producing an exponential distribution of velocities. This
high-velocity warm gas, with speeds up to $\sim 100\kms$, creates a fountain in
the extraplanar regions if the halo potential is too deep for the gas to escape
\citep[see also][]{fielding2018,vijayan2019}.

Regardless of the mechanism for populating high-altitude regions with warm gas,
photoionization from young, massive stars in the disk has long been thought to
be the dominant mechanism that ionizes the Milky Way's DIG 
\citep{bregman1986,reynolds1990, dove1994,
  miller1993, reynolds1995}.

Indeed, past numerical work has shown that photoionization from O and B stars is
capable of ionizing diffuse gas far from the midplane, if such a diffuse gas
layer is present and if there are a sufficient number of low
density paths in the intervening material 
through which ionizing photons may propagate. For example, based
on Monte-Carlo photoionization post-processing of the turbulent hydrodynamic
simulations of \citet{joung2006}, \citet{wood2010} showed that ionizing photons
are able to travel large ($\sim{\rm kpc}$) distances from the midplane and
produce a layer of ionized gas with exponential scale height of $\nelec$ of $500\pc$.
\citet{wood2010} also found that the ionizing photon rate has a strong influence
on the extent and vertical profile of WIM. Similarly, \citet{barnes2014}
post-processed the magnetohydrodynamic (MHD) simulations of \citet{hill2012}, finding that the
additional pressure support from magnetic fields does not significantly change
the high-altitude DIG. An exponential \Halpha{} scale height above $500 \pc$
was found to be $\sim 150 \pc$ and $250 \pc$ for ionizing luminosity per source
of $10^{50} \second^{-1}$ and $10^{49} \second^{-1}$, respectively; this is
insufficient to match the observed extended ionized gas in the Milky Way.
\citet{vandenbroucke2018} repeated the analysis of \citet{barnes2014} for
snapshots from the SILCC simulation of \citet{girichidis2016}, and found that the
exponential scale height of \Halpha\ reached $\sim 600-700\pc$ if cosmic rays
are included, consistent with the observed scale heights of DIG 
in the Milky Way. However, when strong dynamical feedback from supernovae (or cosmic rays) is
absent, as in the radiation hydrodynamic simulations of
\citet{vandenbroucke2019} that included photoionization feedback alone, a DIG
layer at high altitude that reproduces the observations cannot be sustained. 
\rev{ In addition to the above studies, there are a few recent numerical simulations that have included the effect of time dependent 
ionizing radiation feedback in ISM disk models with self-consistent star formation \citep{peters2017, kannan2020}. However,
these simulations have been run for at most 150 Myr (and are thus have not necessarily 
reached a statistically quasi-steady state), and have largely focused on the effect
of early stellar feedback on star formation efficiency and near-midplane structure.}

Massive stars play several roles in the maintenance of the DIG: they
provide the ionizing radiation, and, as supernovae, create the hot and warm
outflows that populate extraplanar regions, while also creating the pathways
that allow ionizing photons to travel far from the midplane. Low density paths
from the major ionizing sources near the midplane are present because the hot
portion of the multiphase ISM (created in supernova shocks) fills a large
fraction of the volume near the midplane \citep{mckee1977,mccray1979}, and
because the warm and cold portions of the ISM are further clumped as a result of
turbulence (which itself is a result of supernova remnant expansion).

The detailed structure of the multiphase ISM is quite sensitive to the
spatio-temporal distribution of supernovae and their correlation with gas
\citep{kim2018}. However, previous simulations of the ISM that have been used as inputs
to radiative transfer models of the WIM
\citep[e.g.][]{joung2006,hill2012,girichidis2016} lack self-consistency in
modeling massive stars and supernovae. The spatial distribution and rate of
supernovae are imposed ``by hand'' for dynamical modeling of the ISM, and the
position and luminosity of ionizing radiation sources are imposed ``by hand''
for radiation post-processing. Imposing SN distributions by hand may affect the
production of high-velocity warm outflows that is responsible for extraplanar gas.
In addition, unrealistic SN distributions may strongly affect the ability of
ionizing photons to propagate long distances through the ISM. For example,
numerical simulations show that if supernova locations are entirely random, the resultant hot volume filling factor is much higher than if all supernovae explode in dense gas
\citep{WalchSILCC2015}. Similarly, non-self-consistent locations of radiation
sources with respect to the gas distribution will also affect photon
propagation.

Because of the multiple roles that massive stars play in 
shaping the structure of the ISM, and the sensitivity of
large-scale ISM structure and dynamics to the spatial correlation between SNe
and gas, it is crucial to model the formation and destruction of massive stars
self-consistently when studying formation of the DIG.
To study gas properties in the extraplanar region, it is also crucial to achieve
uniformly high spatial resolution so that (1) the majority of SN events are
initiated in either the free-expansion or energy-conserving stage and hot
gas is well-resolved when created by shocks, and (2) the interaction at high
altitude between hot winds and warm fountain flows driven by clustered SNe is
properly captured \citep{kim2018,vijayan2019}.

In this work, we use adaptive ray tracing to propagate photons through an MHD
simulation of the star-forming ISM, and investigate the properties of the 
resultant WIM. The approach we use is to post-process snapshots from a model
representative of conditions in the Solar neighborhood, produced within the
Three-phase Interstellar Medium in Galaxies Resolving Evolution with Star
Formation and Supernova Feedback (hereafter TIGRESS) framework \citep{kim2017}.
The TIGRESS framework simulates local patches of a galactic disk at
uniformly
high resolution, including effects of magnetic fields, galactic sheared
rotation, self-gravity, and feedback in the form of FUV heating and supernovae.
In the TIGRESS framework, star cluster formation via local gravitational
collapse and feedback from supernovae are modeled self-consistently. The
distribution of stellar energy sources within the multiphase ISM structure in
the self-consistent TIGRESS framework presumably yields realistic space-time
correlations of radiation sources and absorption sites.

The layout of the paper is as follows. In \autoref{s:method}, we provide details
of the underlying TIGRESS model and the adaptive ray tracing method used to
track UV radiation from massive stars. In \autoref{s:results}, we
review the overall time evolution of the post-processed simulation, the
resulting density structure and statistical properties of the warm ionized gas 
(including gas/dust absorption and escape fractions of radiation),
and construct spatially integrated \Halpha\ line profiles. In \autoref{s:discussion},
we first compare our study with previous numerical work on formation of the WIM,
and then discuss observational applications of our results. Here, we compare our
derived scale heights to observations of both the Milky Way and external
galaxies, and describe our calibration of a clumping correction factor which will allow for
observations of the EM in edge-on galaxies to be converted to a mean electron
density along the line of sight. 

\section{Methods}\label{s:method}

In this section, we describe the MHD simulation used for modeling the
star-forming galactic disk and our procedure for post-processing simulation
snapshots with adaptive ray tracing to compute radiation energy densities and
equilibrium ionization fractions.

\subsection{MHD Simulation}

The TIGRESS framework is built on the grid-based MHD code Athena
\citep{stone2008}, with additional physics modules for shearing box boundary
conditions \citep{stone2010}, 
self-gravity, sink/star particles, and star
formation feedback in the form of clustered and distributed supernovae and
optically thin heating and cooling. \citet{kim2017} present full details of
physical processes modeled and their implementation, results for basic physical
properties of the fiducial Solar neighborhood model, and a numerical convergence
study. Here, we give a brief overview of the numerical methods for star cluster
formation and stellar feedback employed in the TIGRESS framework.


To model the formation of star clusters and their feedback, the TIGRESS
framework employs the sink particle module of \citet{gong2013}, with some
updates. A sink particle, representing a star cluster, is created if the gas in
a cell (1) exceeds the Larson-Penston density threshold at local gas sound
speed, (2) is at a local minimum of the gravitational potential, and (3) has a
converging velocity field in all three directions. The particles' equation of
motion is integrated by a symplectic orbit integration scheme of \citep{quinn2010} 
in the shearing box frame under the total (gas, external, and particle)
gravitational potential. The sink particles accrete mass fluxes into a virtual
control volume ($3^3$ cells surrounding a particle) if gas flows are converging
from all three directions in the particle's rest frame. At the time of particle
formation and whenever a given particle is accreting gas, its control volume is
reset with the extrapolated density, momentum, and energy from the nearby cells,
and only the difference between original and extrapolated values of mass and
momentum is dumped into the sink particle. Sink particles accrete and merge only
before the advent of supernovae.

%
\begin{figure}
  \center{\includegraphics[width=\linewidth]{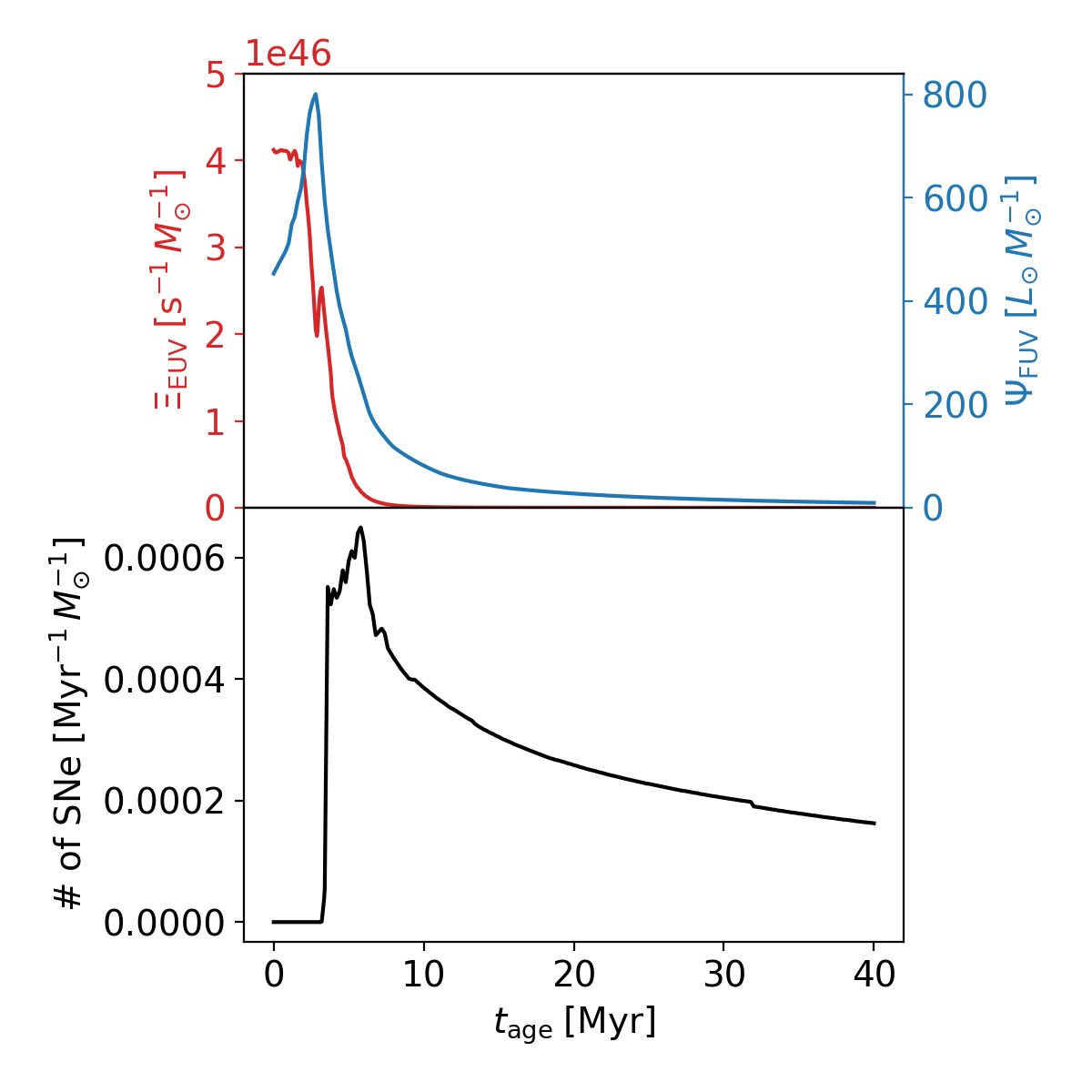}}
  \caption{\textit{Top}: Specific FUV luminosity $\Psi_{\rm FUV}$ (blue) and
    specific ionizing photon rate $\Xi_{\rm EUV}$ (red); \textit{Bottom}:
    Specific SNe rate. Rates are shown as functions of the age of a star cluster
    that fully samples the Kroupa \rev{initial mass function (}IMF). Calculated from 
    STARBURST99.}\label{f:sb99}
\end{figure}

TIGRESS incorporates stellar feedback from young stars in the form of
clustered/distributed SNe as well as FUV radiation. Each star cluster particle
in the simulation represents a star cluster with coeval stellar population
that fully samples the Kroupa \rev{ initial mass function (}IMF) \citep{kroupa2001} with mass-weighted mean age
$t_{\rm sp}$. All star cluster particles with $t_{\rm sp} < 40 \Myr$ can provide
stellar feedback. Depending on the local density of the ambient medium and/or
spatial resolution where a SN event occurs, SN feedback is implemented by either
(1) direct injection of high-velocity SN ejecta (free-expansion stage), (2)
injection of thermal + kinetic energy (energy-conserving, Sedov-Taylor stage),
or (3) momentum (momentum-conserving stage). This ensures that the final radial
momentum added to the surrounding ISM is consistent with the results from
simulations of resolved SN remnant evolution \citep[see][and references
therein]{kim2015}.

The specific FUV luminosity $\Psi_{\rm FUV}$ and SN rate $\xi_{\rm SN}$ of
individual star particles are determined from the STARBURST99 population
synthesis model  \citep[adopting Geneva tracks with zero rotation, Pauldrich model atmosphere, and solar metallicity;][]{leitherer1999}. The blue curve in
the top panel of \autoref{f:sb99} shows $\Psi_{\rm FUV}\equiv L_*/M_*$ of coeval stellar
populations sampling the Kroupa IMF, calculated from STARBURST99.

The clustered SNe occur at the positions of star particles with age
$t_{\rm sp} \gtrsim 3.5 \Myr$ (bottom panel of \autoref{f:sb99}), 
while the distributed SNe
are modeled via runaway OB star particles that are ejected from star cluster
particles with an ejection velocity distribution consistent with a binary
population synthesis model \citep{eldridge2011}. The clustered and distributed
SNe constitute 2/3 and 1/3 of the total SNe events, respectively.

The heating for cold and warm gas ($T < 2\times 10^4\Kel$, see
\autoref{tab:phases}) represents dust photoelectric heating caused by FUV
photons \citep{wolfire1995}. The local FUV intensity is assumed to be
proportional to the total FUV luminosity of feedback particles (with a
correction for dust shielding expected for a plane-parallel slab as described in
\citealt{ostriker2010}), but is spatially uniform across the simulation domain.
An optically-thin cooling function following \citet{koyama2002} is adopted for
cold and warm gas ($T \lesssim 10^{4.2} \Kel$) and that of
\citet{sutherland1993} is used for hot gas under the assumption of collisional
ionization equilibrium.

\begin{table}[]
    \centering
    \caption{Temperature boundaries for thermal phases}
    \begin{tabular}{l|c}
    \toprule
    Phase & Temperature boundary \\
    \midrule
         Cold &  $T<184\Kel$ \\
         Unstable &  $184\Kel<T<5050 \Kel$ \\
         Warm &  $5050 \Kel< T <2\times 10^4 \Kel$ \\
         Hot\tablenotemark{a} & $T>2\times10^4 \Kel$ \\
    \bottomrule
    \end{tabular}
    \tablenotetext{a}{The hot phase described above includes both the ionized
      and hot phases in \cite{kim2018}.}
    \label{tab:phases}
\end{table}

We use the Solar neighborhood model (R8) presented in Kim et al. (2020, in
prep), which adopts the same galactic conditions analyzed in \citet{kim2017},
\citet{kim2018}, and \citet{vijayan2019}, with additional updates for the
treatment of sink particle accretion as described above. We adopt the
galactocentric distance $R_0 = 8\kpc$, angular velocity of local galactic
rotation $\Omega = 28 \kms\kpc^{-1}$, and shear parameter
$d\ln\Omega/d\ln R = -1$. The initial gas surface density is
$\Sigma_0 = 13 \Msun\pc^{-2}$. The total gas mass decreases gradually as gas
turns into stars and outflows escape the vertical boundaries. The box size for
this model is $L_x = L_y = 1024 \pc$ and $L_z = 7168\pc$ with a uniform grid
spacing $\Delta x = 4\pc$. The simulation is run for $660 \Myr$ (3 orbital
times), long enough for the system to establish a statistically quasi-steady
state in which the physical state of the multiphase, turbulent ISM is
consistently generated from star formation and supernova plus heating feedback.
The impact of the initial transient evolution is minimal after $\sim 100 \Myr$. Full data
output snapshots are taken at intervals of $0.97\Myr(=1\pc/(1\kms))$.

\subsection{Post-processing with Adaptive Ray Tracing}

The MHD simulation snapshots are post-processed with the adaptive ray tracing
algorithm implemented in the \textit{Athena} code by \citet{kimjg2017}, which
solves the equation of radiative transfer for systems containing multiple point
sources (neglecting scattering). For each snapshot, we read in the hydrogen
number density ($\nH$) and temperature ($T$) of gas, star particle data (position,
mass, age for each source), and the simulation time ($t$) as inputs.

For a sink particle of mass $M_{\rm sp}$ and mass-weighted mean age
$t_{\rm sp} (< 40 \Myr)$, we calculate the ionizing photon production rate as
$Q_{\rm i,sp} = \Xi_{\rm EUV}(t_{\rm sp})M_{\rm sp}$, where $\Xi_{\rm EUV}$ is the
ionizing photon production rate per unit stellar mass (red line in
\autoref{f:sb99}(a)). Because massive stars with lifetimes of $\sim 3\Myr$
dominate the ionizing photon output, $\Xi_{\rm i}$ is roughly constant at
$4 \times 10^{46} \second^{-1} \Msun^{-1}$ before the onset of the first SN
($t_{\rm sp} \lesssim 3.5\Myr$), and declines sharply afterwards. Similarly, the
FUV luminosity of each star particle is calculated as
$L_{\rm FUV,sp} = \Psi_{\rm FUV}(t_{\rm sp})M_{\rm sp}$.

The adaptive ray tracing injects photon packets at the position of each sink
particle and carries them along rays, calculating the local optical depth, the
corresponding photon absorption rate by gas and dust, and the radiation energy
density in two frequency bins (EUV and FUV). The ray direction is
determined using the HEALPix scheme \citep{gorski2005}, which subdivides
the unit sphere into $12\times 4^{\ell}$ equal-area pixels at HEALPix level
$\ell$. We adopt the initial HEALPix level $\ell_0 = 4$ for the injected photon
packets. Rays are split to ensure that each grid cell is sampled by at least
four rays per source (unless one of the termination conditions is met; see
below).

Similar to other flow attributes, the shearing-periodic boundary conditions are
applied to rays crossing the radial ($x$) boundaries. For example, if a ray
exits the far radial boundary at $y=y_0$, it re-enters the near radial boundary
at $y = y_0 + {\rm sgn}(L_y - \Delta y - y_0) \times (L_y - \Delta y)$, where
$\Delta y = (q\Omega L_x t \mod L_y)$ is the shear displacement in the
$y$-direction, with the position of the source offset accordingly. The azimuthal
($y$) boundary condition is strictly periodic.

Each photon packet is followed along a ray until one of the following conditions
is satisfied: (1) the ray needs to split; (2) the optical depth from the source
is greater than 10; (3) the ray exits the computational domain in the
\rev{$d_{\rm xy}$ is greater than
$d_{\rm xy, \rm max}= L_x = 1024 \pc$}.\footnote{\rev{While this condition
    does not remove photon packets based on the vertical distance traveled, a
    photon packet (injected near the midplane) is terminated before reaching the
    vertical boundary if the angle between the ray direction and $z$-axis is
    greater than
    ${\rm cos^{-1}}\left(1/\sqrt{1 + (2d_{\rm xy, \rm max}/L_z)^2}\right)
    \approx 30^{\circ}$.}
} Without the last condition, a small fraction of photon packets traveling a
very long distance ($\gg
L_{x}$) on horizontal optically-thin rays make the computational cost of ray
tracing expensive.
We have checked that the use of larger $d_{\rm xy, \rm
  max}$ has little impact on the outcome of our analysis.


For the opacity per unit length, we take
$\chi_{\rm i} = \nHI \sigma_{\rm ph} + \nH \sigma_{\rm d,i}$ and
$\chi_{\rm n} = \nH \sigma_{\rm d,n}$ for ionizing and non-ionizing radiation,
respectively, where $\sigma_{\rm ph} = 2.7 \times 10^{-18} \cm^2\,{\rm H}^{-1}$
is the frequency-averaged photoionization cross section and
$\sigma_{\rm d,i} = \sigma_{\rm d,n} = 1.17 \times 10^{-21} \cm^2\,{\rm H}^{-1}$
is the frequency-averaged dust absorption cross section
\citep[e.g.,][]{draine2011b}.

\subsection{Runaways}\label{s:runaways}

In the TIGRESS framework, the age and orbits of runaway star particles are
tracked consistently, and the ionizing photon rate of individual runaway star
particles can be inferred from the total SN rate (bottom panel in
\autoref{f:sb99}) and mass-luminosity relation for main-sequence stars
\citep{parravano2003}. To examine the effect of runaways on the ionization of
gas, we post-processed the simulation snapshots including both clusters and
runaway particles as ionizing sources. We find that including runaways does not
change the extent and distribution of WIM significantly, because they contribute
insignificantly to the total ionizing photon budget; the details of this
calculation are summarized in \autoref{appendix}.

\subsection{Ionization State Calculation}

After the completion of a ray trace, we calculate the ionization state of a gas
cell assuming simple ionization-recombination balance
$\mathcal{I}_{\rm phot} + \mathcal{I}_{\rm coll} = \mathcal{R}$, where
\begin{align}
  \mathcal{I}_{\rm phot} &= \nHI\sigma_{\rm ph} c
                           \mathcal{E}_{\rm i}/(h\nu_{\rm i}) \\
  \mathcal{I}_{\rm coll} & = \gamma_{\rm coll}\nHI\nelec \\
  \mathcal{R} & = \alphaB \nHII \nelec
\end{align}
are the local photoionization, collisional ionization, and radiative
recombination rates, respectively. Here, $\mathcal{E}_{\rm i}$ is the radiation
energy density for ionizing radiation, $h\nu_{\rm i} = 18 \eV$ the mean energy
of ionizing photons,
$\alphaB = 2.59 \times 10^{-13} (T/10^4\Kel)^{-0.7}\cm^3\second^{-1}$ the case B
recombination coefficient \citep{krumholz2007}\footnote{In their Monte-Carlo
  photoionization simulation, \citet{barnes2014} found that the majority of
  diffuse ionizing photons resulting from the recombination to the ground state
  are re-absorbed in-situ, suggesting that the on-the-spot approximation is
  reasonable.},
$\gamma_{\rm coll} = 5.84 \times 10^{-11} \sqrt{T/{\rm K}}\exp({{-157821\,{\rm
      K}}/{T}}) \cm^3\second^{-1}$ the collisional ionization rate coefficient
\citep{tenorio-tagle1986}, $c$ the speed of light, $\xn = \nHI/\nH$ the neutral
fraction, and $\nelec = \nHII = (1 - \xn)\nH$ the free electron number density.
Note that for simplicity, we neglect the ionization of helium and other species,
and free electrons released by them. Solving for $\xn$ gives the equilibrium
neutral fraction as
\begin{equation}
  \xn = \dfrac{2\alphaB\nH}{(\Gamma + (2\alphaB +
    \gamma_{\rm coll})\nH) + \sqrt{(\Gamma + \gamma_{\rm coll}\nH)^2 +
      4\Gamma\alphaB\nH}}\,,
\end{equation}
where $\Gamma = \mathcal{I}/\nHI$  \citep[e.g.,][]{altay2013}. In the
absence of photoionization ($\Gamma=0)$,
$x_{\rm n,coll} = \alphaB / (\alphaB + \gamma_{\rm coll})$.

In addition to the ionization balance, we assume that the thermal balance
between heating and cooling keeps the temperature of photoionized gas at a
constant value $T_{\rm ion} = 10^4 \Kel$. We alter the temperature of gas cells
exposed to ionizing radiation as
$T = T_{\rm ion} - (T_{\rm ion} - T_0){x_{\rm n,eq}}/{(2.0 - x_{\rm n,eq})}$ if
$T_0 < T_{\rm ion}$, where $T_0$ is the temperature of gas in the MHD
simulation. By doing so, the temperature of (collisionally ionized) hot gas
remains unchanged, and the temperature of photoionized gas becomes
$T \approx T_{\rm ion} = 10^4 \Kel$.

Since the change in gas temperature affects the recombination rate and the
Str\"{o}mgren volume calculation, the whole procedure (ray trace + equilibrium
neutral fraction) is repeated until (1) the total volume of ionized gas
converges to within $0.01\%$, and (2) the total ionization rate balances the
recombination rate to within $0.01\%$. Each snapshot requires 20--30 iterations
to converge to the desired accuracy.

We divide gas into four different phases based on temperature:
$T > 2 \times 10^4 \Kel$ for hot, $5050 \Kel < T < 2 \times 10^4 \Kel$ for warm,
$184 \Kel < T < 5050 \Kel$ for unstable, and $T \le 184 \Kel$ for cold (see
\autoref{tab:phases}). As noted in \autoref{s:intro}, we use ``warm ionized
gas'' (or WIM) as a generic term to refer to both dense and diffuse photoionized
gas in the simulation, and do not make a strict distinction between dense
ionized gas (at low altitudes) and low-density ionized gas (both at low- and
high-altitudes) since the dynamical expansion of ``classical \HII\ regions'' is
not modeled in the MHD simulation. Instead, we characterize the properties of
warm ionized gas at varying densities and distance from the midplane\rev{,
  defined as $z=0$},\footnote{\rev{Although we 
  fix the position of the ``midplane''
    for this analysis to $z=0$, the center of mass height of the warm gas has a
    median position of $z=2$ pc, and a 25\thh{} (75\thh{}) percentile value of
    -36 pc (58 pc). The cold gas has a median center of mass height of $z=5$ pc,
    and a 25\thh{} (75\thh{}) percentile height of -22 (27) pc.}}. For practical
purposes, we adopt $|z| = 200 \pc$ as a dividing line between low and high
altitude ionized gas and regard all of the warm ionized gas above $|z| = 200\pc$
as DIG.

\begin{figure*}[t!]
  \center{\includegraphics[width=\linewidth]{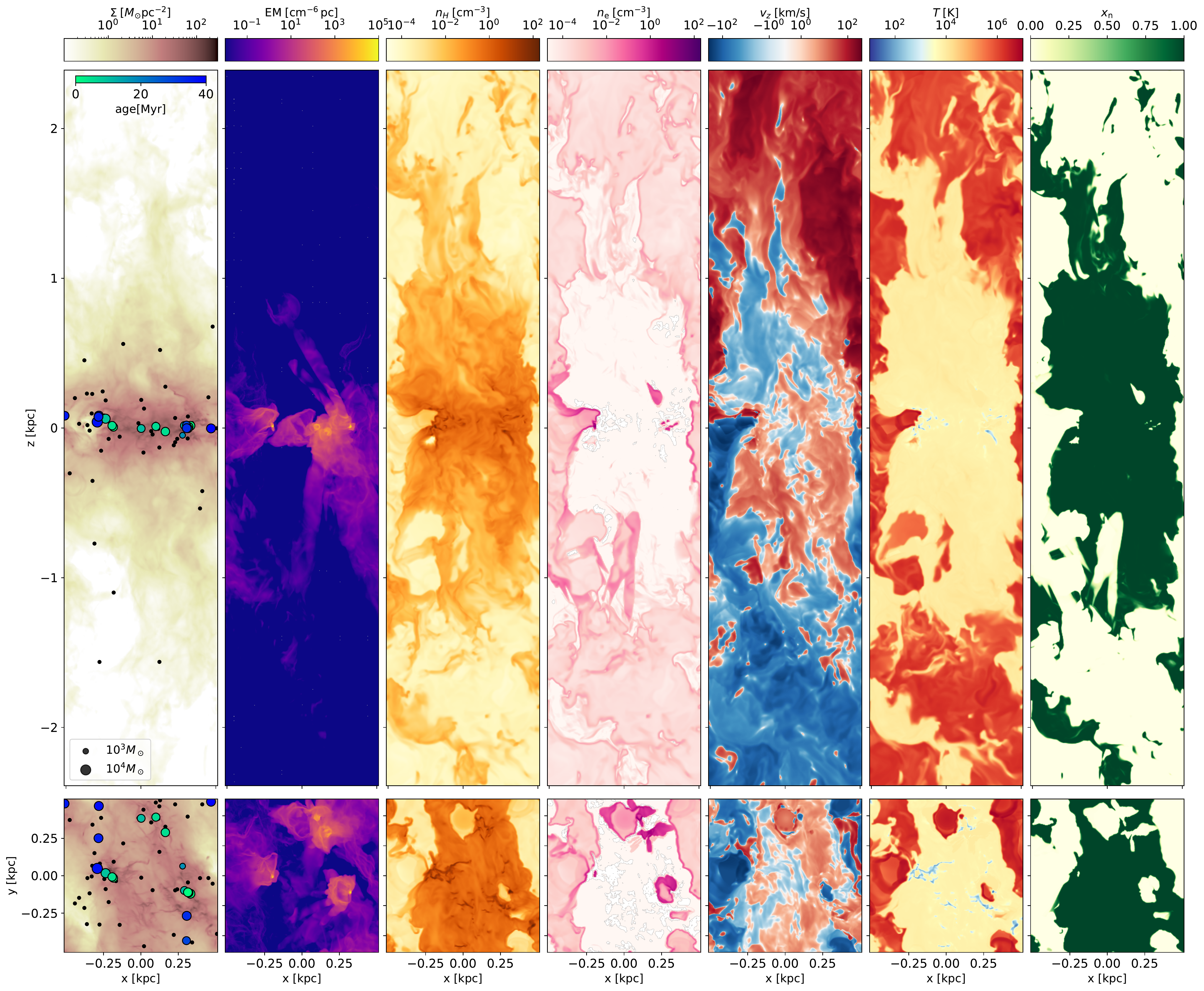}}
  \caption{A sample post-processed snapshot at $t=551\Myr$ with total ionizing
    photon rate $\Qi = 8.0 \times \times 10^{50} \second^{-1}$. The far left
    panels show the gas surface density projected along the $y$- (top) and
    $z$-directions (bottom). The projected positions of the star/sink particles
    are shown as colored circles, with size and color indicating mass and
    mass-weighted age ($t_{\rm sp}$), respectively. Runaway stars are shown as
    black dots. Continuing to the right, the panels show the EM (integrated
    electron density squared, ${\rm EM} = \int \nelec^2 d\ell$), slices through
    the center of the simulations box of hydrogen number density $\nH$, electron
    number density $\nelec$, vertical velocity $v_z$, gas temperature $T$, and
    neutral fraction $\xn = \nHI/\nH$. The full vertical extent of the
    simulation domain is $-3.584\kpc < z < 3.584\kpc$. }\label{f:snapshot1}
\end{figure*}

\begin{figure*}[t!]
  \center{\includegraphics[width=\linewidth]{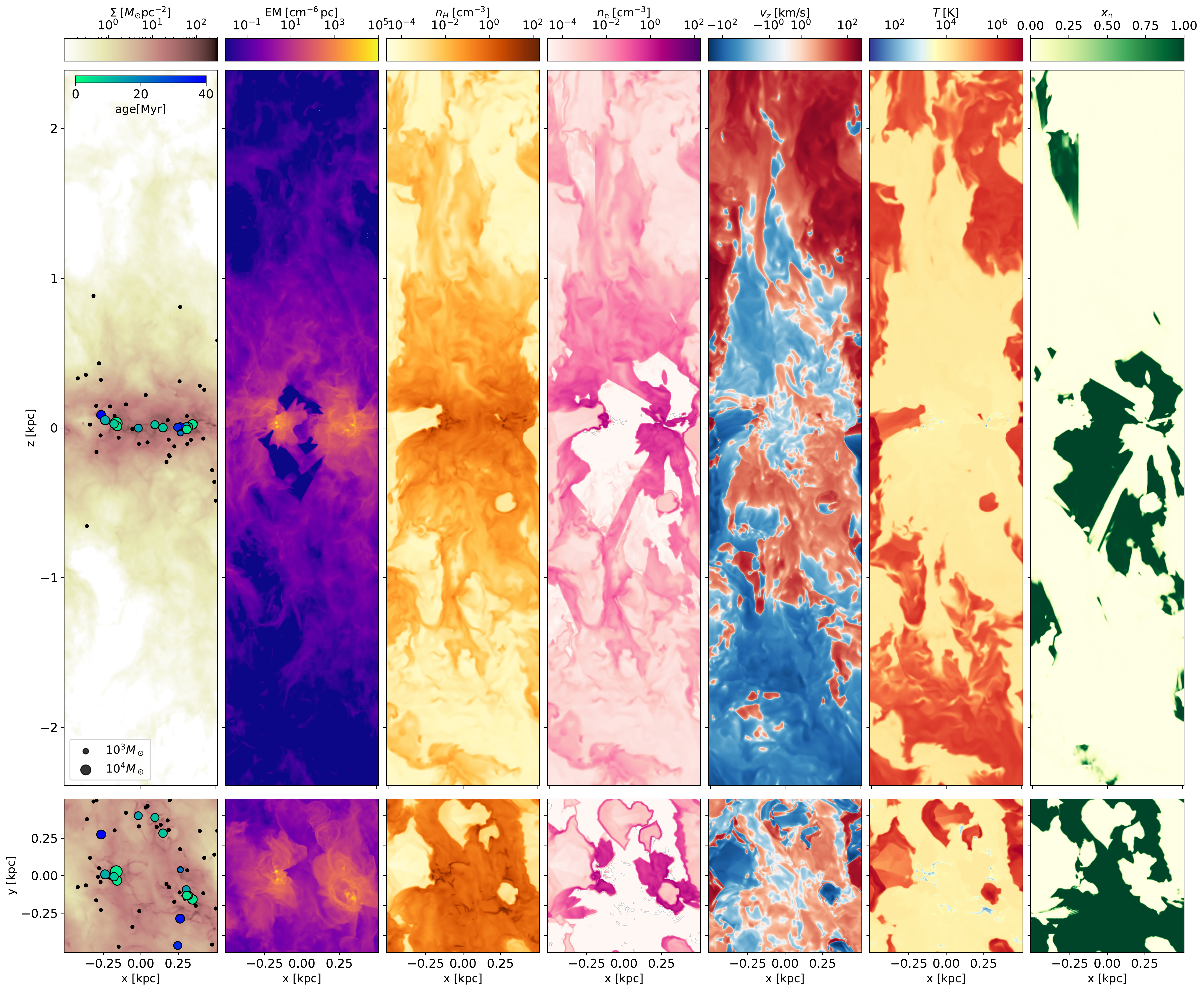}}
  \caption{Same as \autoref{f:snapshot1}, but at $t=555\Myr$ when
    $\Qi = 1.1 \times 10^{51}\second^{-1}$. With only $4\Myr$ time difference
    from the snapshot in \autoref{f:snapshot1}, the density, temperature, and
    velocity structure has changed little. However, the WIM layer is
    significantly more extended than in \autoref{f:snapshot1}, as can be seen in
    the projected electron density (EM, third panel from left) and electron
    density slice (fourth panel from right), and neutral fraction (rightmost
    panel).}\label{f:snapshot2}
\end{figure*}

\begin{figure*}[t!]
  \center{\includegraphics[width=\linewidth]{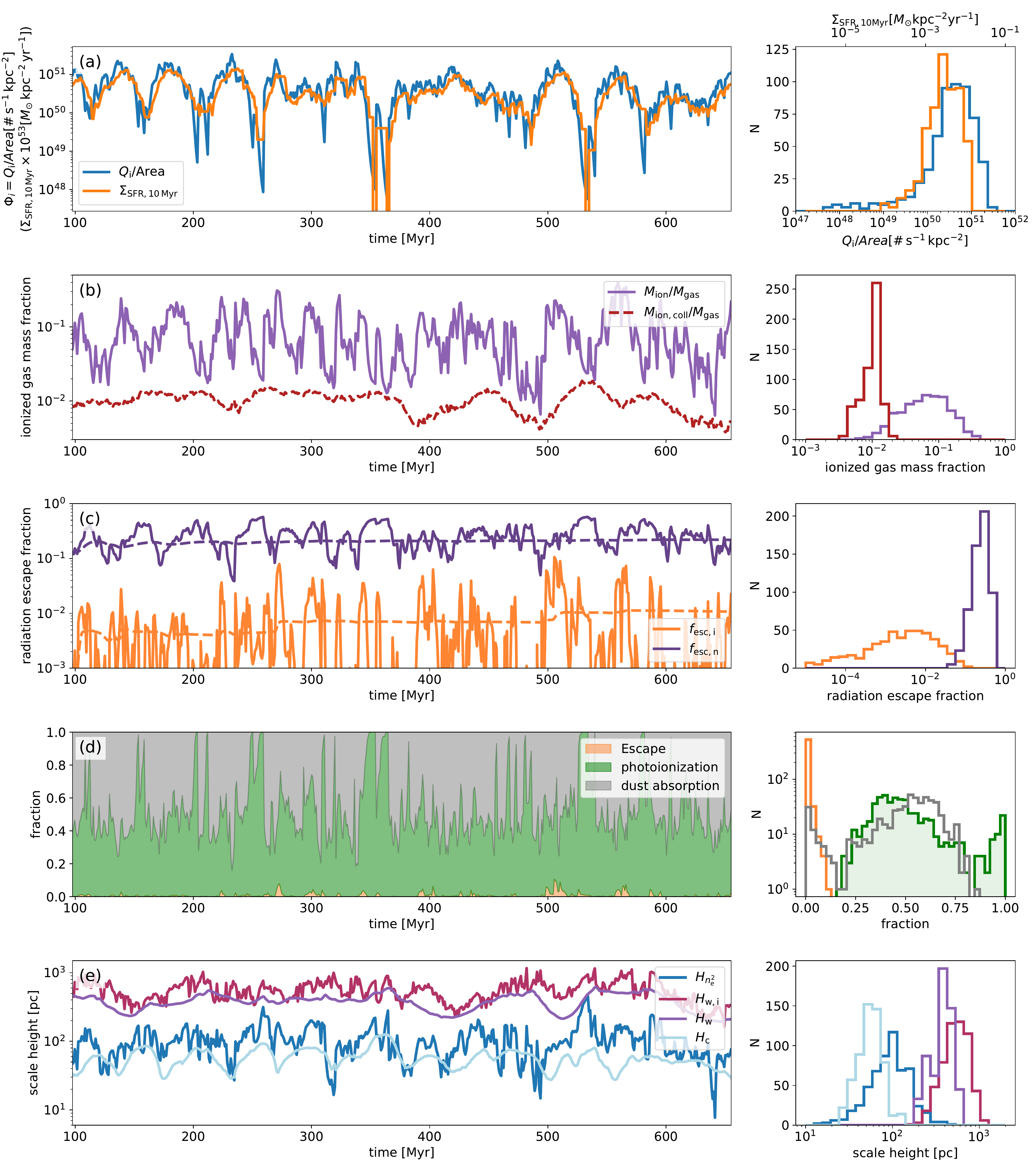}}
  \caption{The time evolution (left) and distribution over time (right) of
    various box-integrated quantities. \textit{Row \textsf{(a)}:} Ionizing
    photon rate per unit area ($\Phi_{\rm i}=\Qi/(L_x L_y)$, in photon s$^{-1}$
    kpc$^{-2}$) is shown in blue. The $10 \Myr$ averaged star formation rate
    surface density is shown in orange. \textit{Row \textsf{(b)}:} The fraction
    of ionized (photoionization + collisional ionization) mass is shown by the
    solid curve (violet), while the fraction of collisionally ionized mass is
    shown by the dashed curve (crimson). \textit{Row \textsf{(c)}:} The escape
    fractions of ionizing (orange) and non-ionizing (purple) radiation. The
    solid and dashed lines plot instantaneous and cumulative escape fractions,
    respectively. \textit{Row \textsf{(d)}:} The fraction of ionizing photons
    that escape from the box (orange), photoionize neutral hydrogen (green), and
    are absorbed by dust (grey). \textit{Row \textsf{(e)}:} Scales heights,
    defined as rms distance from the midplane, of various components
    (\autoref{e:scaleheight}). Shown are warm ionized gas (magenta), total warm
    gas (purple), and cold gas (light blue), as well as $\nesq \propto {\rm EM}$
    (dark blue). }\label{f:hst}
\end{figure*}

%
\begin{figure*}[t!]
  \center{\includegraphics[width=\linewidth]{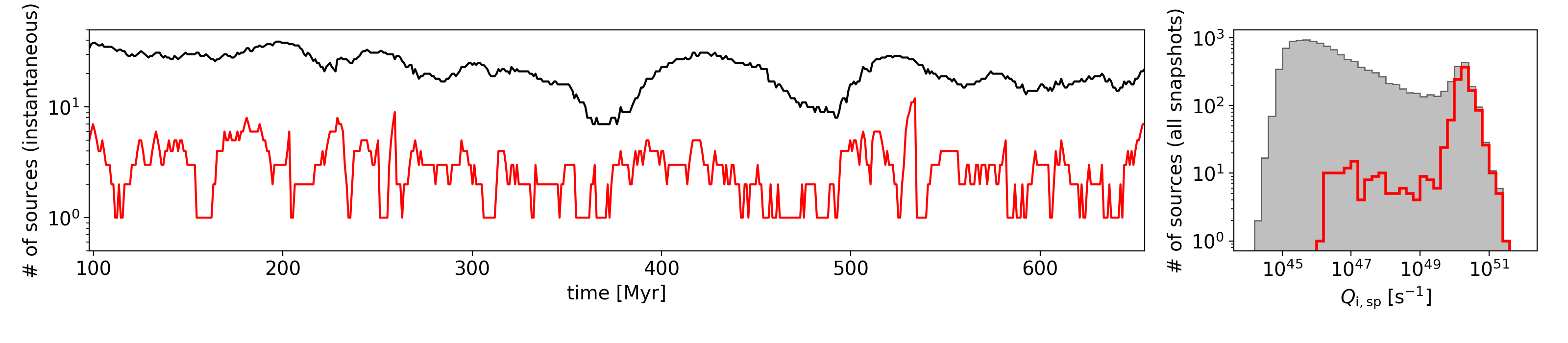}}
  \caption{\textit{Left:} total number of ionizing source particles (black) and
    the number luminous source particles needed to account for $>90\%$ of the
    total ionizing photon rate (red) as functions of time. \textit{Right:}
    distribution of the ionizing photon rate of source particles $Q_{\rm i,sp}$
    for all snapshots. As in the left panel, the grey histogram shows the
    distribution for all sources, while the red unfilled histogram shows the
    distribution for the brightest clusters that are responsible for $>90\%$ of
    $\Qi$. Although we include all star particles with $t_{\rm sp} < 40 \Myr$ as
    active ionizing sources for a given snapshot, the total ionizing photon
    budget is dominated by a few young clusters.}\label{f:numsrc}
\end{figure*}

\section{Results}\label{s:results}

Here we present the results of the post-processing described above. We first
examine the overall evolution of the simulation and tracked quantities therein,
including statistics for radiation sources and sinks. We then analyze properties
of warm ionized gas, including vertical profiles, volume filling factor, scale
heights, and various statistical measures of the EM distribution. We also
present synthetic \Halpha\ line profiles, commenting on comparisons to the WHAM
\citep[Wisconsin \Halpha\ Mapper, ][]{reynolds1998} survey and prospects for
identifying DIG based on spatially integrated velocity information. Comparison of our results to
observations of gas scale heights in the Milky Way and external galaxies will be
made in \autoref{s:discussion}.

\subsection{Overall Evolution}\label{s:evolution}

After $\sim 100\Myr$, the MHD simulation reaches a quasi-steady state in which
the star formation rate (SFR) is self-regulated and ISM phases are in balance.
The energy input from newly formed stars stirs turbulent motions and heats the
ISM, maintaining the (turbulent + magnetic + thermal) pressure to offset the
vertical weight of the ISM disk and hence to prevent runaway gravitational
collapse. Meanwhile, expansion of hot superbubbles created by clustered SNe
drives multiphase outflows consisting of hot winds and warm fountains.  While the
hot phase outflow achieves high enough velocity to escape into the galaxy's
halo, most of the warm phase outflow has $|v_z|<100\kms$ and is unable to
overcome the large-scale gravitational potential, falling back onto the disk
eventually. Averaged over several star-formation cycles, the properties of the
hot wind and warm fountain are in a statistically steady state (see
\citealt{kim2017,kim2018,vijayan2019} for more quantitative analyses of star
formation rate, thermal phase balance, and outflow properties).

To illustrate the structure of the ISM in the post-processed TIGRESS
simulations, \autoref{f:snapshot1} and \autoref{f:snapshot2} show two example
snapshots at $t= 551 \Myr$ and $555 \Myr$. Each figure shows, from left to
right, projections of the gas density ($\Sigma = \int \rho d\ell$, overlaid with
star particle positions), emission measure (${\rm EM} = \int \nesq d\ell$),
slices of hydrogen number density ($\nH$), free electron number density
($\nelec$), gas velocity in the vertical direction ($v_z$), gas temperature
($T$), and neutral fraction ($\xn$). The projections are along the $y$-axis
(equivalent to the azimuthal direction; top panels) or $z$-axis (vertical
direction; bottom panels), while slices are through $y=0$ (top) or $z=0$
(bottom).\footnote{We note that with $L_z = 7168 \pc$ the projections along the
  $z$-axis are comparable to the face-on view of galaxies. Quantities projected
  along $y$-axis depend on the horizontal box size  $L_y=1024 \pc$ adopted for the
  simulation, but modulo rescaling for relative path length provide a view of
  the ISM similar to that of an edge-on galaxy.}

To provide a sense of the physical scope and the dynamic range of our model,
\autoref{f:hst} shows the time evolution (left) and distributions over time
(right) of various global quantities. \autoref{tab:summarystats} provides a summary
of statistical properties for these quantities: ionizing photon rate, star
formation rate, mass fraction of ionized gas, escape fraction of ionizing and
non-ionizing photons, fraction of photons lost to photoionization and dust
absorption, and scale heights of the various gas components.

As shown qualitatively in \autoref{f:snapshot1} and \autoref{f:snapshot2}, and
quantitatively in \autoref{f:hst}, both the structure and the extent of the WIM
are highly variable in space and time. This variability is driven by the
variability of $\Qi$ and the presence or absence of escape channels for ionizing
photons surrounding ionizing sources (see \autoref{s:Qi} and \autoref{s:fesc}).
For example, the distributions at $t = 555 \Myr$ in \autoref{f:snapshot2} shows
both inflowing and outflowing regions with a significant extraplanar DIG layer
extending over $1\kpc$ from the midplane. By contrast, in the snapshot at
$t=551 \Myr$ (\autoref{f:snapshot1}) there is relatively little warm ionized gas
far from the midplane (based on lower EM at large $|z|$), despite these
snapshots being separated by just $\sim 4\Myr$, possessing a similar ionizing
photon rate, and having a similar outflow rate of warm gas. This demonstrates
the importance of radiative transfer to the DIG, which adds to the significant
time variability already demonstrated for the ISM flow in the TIGRESS
simulation. In spite of the time variability, we find that the existence of an
extended WIM profile is commonplace.

We note that while the slices of $\xn$ show that gas is either fully neutral or
fully ionized because we evolve to equilibrium, at low density the recombination
rate is low enough that in reality gas may remain partially-ionized even when
not directly exposed to radiation \citep[e.g.,][]{dong2011}. Evaluation of the
importance of this effect for enhancing the DIG will require inclusion of
ray-tracing and ionization/recombination in future time-dependent simulations.

\subsubsection{Star formation, ionizing photon budget, and source
  properties}\label{s:Qi}

Row \textsf{(a)} in \autoref{f:hst} shows that the SFR per unit area
$\Sigma_{\rm SFR,10\Myr}$ (orange, calculated from the mass of star particles
with $t_{\rm sp} < 10 \Myr$) exhibits significant temporal fluctuations. The
resulting ionizing photon production rate per unit area
$\Phi_{\rm i} \equiv \Qi/(L_x L_y)$\footnote{While $\Phi_{\rm i}$ is usually
  reported in cgs units in the literature, we adopt a unit
  ${\#}\, {\rm s}^{-1}\,{\rm kpc}^{-2}$ that connects more intuitively to
  $\Sigma_{\rm SFR}$. Note that
  $10^{50}\,{\rm s}^{-1}\,\kpc^{-2} = 1.05 \times 10^7\,{\rm cm}^{-2}\,{\rm
    s}^{-1}$.} (blue) is well correlated with
$\Sigma_{\rm SFR,10\Myr}$\footnote{Overall, $\Sigma_{\rm SFR,10 Myr}$ lags
  slightly behind $\Phi_{\rm i}$ because the timescale on which SFR is measured
  is longer than the characteristic lifetime of ionizing stars. The EUV-weighted
  mean age of star clusters in the simulation is 2.1 Myr, as compared to the 10
  Myr timescale over which SFR is averaged.}, with a more pronounced fluctuation
amplitude. The typical ionizing photon rate per unit area is
$\Phi_{\rm i} = \Qi/(L_x L_y) = 4.2_{-2.3}^{+4.0} \times 10^7$ s$^{-1}$
cm$^{-2}$ ($4.0_{-2.2}^{+3.8}\times 10^{50}$ s$^{-1}$ kpc$^{-2}$), which is
roughly consistent with the observational estimate
$5.2 \times 10^7 \second^{-1}\,\cm^{-2}$ in the solar neighborhood
(\citealt{mckee1997}; see also \citealt{abbott1982, dove1994, vacca1996}). An
overview of the summary statistics for the quantities shown in \autoref{f:hst}
is given in \autoref{tab:summarystats}.

One reason for the strong variability in the DIG is that a small number of
sources are responsible for most of the ionization, and as a result localized
absorption near the midplane can cast large volumes at high latitude into
ionization ``shadows.'' The left panel of \autoref{f:numsrc} shows the time
evolution of the number of active ionizing sources (black) and the minimum
number of ionizing sources to account for 90\% of the total ionizing photon rate
(red). While there are 22 active ionizing sources on average, the majority of
the ionizing photon budget is supplied by just a few ($\sim 3$) luminous sources
with $t_{\rm sp} \lesssim 5\Myr$. The right panel of \autoref{f:numsrc} shows
the distributions over time of the ionizing photon rate of all individual
sources (grey) and of the sources that make up $>90\%$ of the total ionizing
photon rate (red), which indicates that young star clusters with
$Q_{\rm i,sp} \sim 10^{50}\,\second^{-1}$ dominate the ionizing photon
production. The cutoff at $Q_{\rm i, sp} \sim 10^{45}\second^{-1}$ in the grey
histogram is caused by the choice $40 \Myr$ as the maximum age of ionizing
sources.

\subsubsection{Mass of Ionized Gas}\label{s:massion}

Row \textsf{(b)} of \autoref{f:hst} (crimson line) shows that the mass of
collisionally ionized gas that would be produced by supernova shocks in the
absence of photoionization in the MHD simulation (i.e.,
$\int \nH\gamma_{\rm coll}/(\gamma_{\rm coll} + \alphaB) dV$) is only $\sim 1\%$
of the total gas mass. Including photoionization in the post-processing boosts
the ionized mass fraction significantly (violet line), unless there is little
recent star formation. The ionized gas mass fraction shown in \autoref{f:hst}
includes both dense and diffuse components. We find that the low-density DIG at
$|z| > 200\pc$ is substantial in mass, but not in emission (see also
\autoref{s:nedist}), because the emissivity is proportional to $\nesq$. The mass
of DIG at $|z| > 200\pc$ represents $\sim 7\%$--$76\%$ of the total ionized gas
mass with a mean of $40\%$. In contrast, the fraction of the total \Halpha\
emission originating from DIG at $|z| > 200\pc$ is minor, ranging from
$5.6\times 10^{-5}$ to $0.36$ with a mean of $0.05$.

\subsubsection{Gas Scale Heights}\label{s:H}

Row \textsf{(e)} of \autoref{f:hst} shows the time evolution and distributions
of the scale heights of warm ($H_{\rm w}$), warm ionized ($H_{\rm w,i}$), and
cold ($H_{\rm c}$) gas. In addition, we show the scale height of emissivity from
warm ionized gas $\propto n_e^2$ ($H_{\nesq}$). The two-phase (cold + warm) scale
height (not shown) is approximately equal to the total warm gas scale height.
  
The reported scale heights are defined as
\begin{equation}\label{e:scaleheight}
  H_{q} \equiv \sqrt{\frac{\int \langle q\rangle z^2 dz}{\int \langle q\rangle
      dz}}
\end{equation}
where the brackets $\langle \cdot \rangle$ denotes the horizontal average, and
the $q$ is the quantity over which the scale height is computed (e.g. the number
density ($\nH$), free electron density ($\nelec$), or the square of the electron
density ($\nelec^2$) of a cold or warm phase).\footnote{For (Gaussian,
  exponential, ${\rm sech}^2$) vertical profiles with
  $q \propto (e^{-z^2/2h^2}$, $e^{-|z|/h}$, ${\rm sech}^2(z/h))$, produce scale
  heights of $H_q = (1, \sqrt{2}, 0.91)h$}

The median scale height of (star-forming) cold gas is only $55 \pc$, which is roughly 
comparable to the $\Qi$-weighted scale height of the ionizing sources
($38 \pc$). Both are in turn roughly comparable to the Gaussian scale height of
O-B5 stars of $63 \pc$ measured by Hipparcos \citep{maizapellaniz2001}. The warm
gas scale height fluctuates between about $250 \pc$ and $500 \pc$, with a median
value of $393 \pc$. The scale height of warm ionized gas is somewhat larger,
with a median value of $540 \pc$. The median value of $H_{\nesq}$ is only
$94 \pc$, considerably smaller than $H_{\rm w,i}$, and only slightly larger than
$H_c$. This suggests that the dense ionized gas near ionizing sources
constitutes a major fraction of the total emission from warm ionized gas (see
\autoref{s:nedist}). For this reason, the scale heights of WIM ($H_{\rm w,i}$ and
$H_{\nesq}$) derived from \autoref{e:scaleheight} can be quite different from
observational scale heights derived from different approaches (e.g., by fitting
an exponential to the vertical component of EM or DM, excluding dense ionized
gas). We defer detailed the analysis of WIM scale height and its comparison to
observations until \autoref{s:scaleheight}.



\begin{deluxetable*}{cccccccccccc}
\tablecaption{Summary Statistics of Selected Properties}
\tablewidth{0pt}
\tablehead{
\colhead{} & 
\colhead{$\Sigma_{\rm SFR,10Myr}$} &
\colhead{$\Phi_{\rm i}$} &
\colhead{$M_{\rm ion}/M_{\rm gas}$} &
\colhead{$f_{\rm esc,i}$} &
\colhead{$f_{\rm dust,i}$} &
\colhead{$f_{\rm esc,n}$} &
\colhead{$H_{\rm c}$} &
\colhead{$H_{\rm w}$} &
\colhead{$H_{\rm w,i}$} &
\colhead{$H_{\nesq}$} &
\\
\colhead{} &
\colhead{$\left(\dfrac{10^{-3}M_{\odot}}{\kpc^2\,{\rm yr}}\right)$} &
\colhead{($10^{50}{\rm s}^{-1}\,{\rm kpc}^{-2}$)} &
\colhead{(\%)} &
\colhead{(\%)} &
\colhead{(\%)} &
\colhead{(\%)} &
\colhead{(pc)} &
\colhead{(pc)} &
\colhead{(pc)} &
\colhead{(pc)} &
\\
\colhead{(1)} & 
\colhead{(2)} &
\colhead{(3)} &
\colhead{(4)} &
\colhead{(5)} &
\colhead{(6)} &
\colhead{(7)} &
\colhead{(8)} &
\colhead{(9)} &
\colhead{(10)} &
\colhead{(11)} &
}
\startdata
25\thh{} & 1.51 & 1.82 & 3.38 & 0.02 & 40.0 & 17.2 & 44.4 & 301 & 423 & 65.0 \\
50\thh{} & 2.98 & 3.99 & 6.30 & 0.19 & 53.0 & 24.8 & 55.4 & 393 & 541 & 94.2 \\
75\thh{} & 5.84 & 7.79 & 10.2 & 0.82 & 61.4 & 33.9 & 68.0 & 460 & 680 & 134 \\
Mean     & 4.00 & 5.54 & 8.29 & 1.08\tablenotemark{a} & 57.1\tablenotemark{a} & 21.8\tablenotemark{a} & 57.5 & 384 & 557 & 105\\
\enddata
\tablenotetext{a}{The reported values are $Q_{\rm i}$-averaged (or cumulative)
mean, for example, $\int (f_{\rm esc,i}Q_{\rm i})dt/\int Q_{\rm i} dt$.}
\tablecomments{Column (1) Summary statistic (percentile or mean) Column (2) Star formation rate
per unit area averaged over $10\Myr$. Column (3) mass fraction of ionized gas.
Column (4) Ionizing photon rate per unit area. Column (5) Escape fraction of
ionizing radiation. Column (6) Dust absorption fraction of ionizing radiation.
Column (7) Escape fraction of non-ionizing radiation. Columns (8)--(11) Scale
height of cold, warm, warm ionized, and $\nesq$ (see \autoref{e:scaleheight}).
}\label{tab:summarystats}
\end{deluxetable*}

\subsection{Escape fraction and dust absorption fraction}\label{s:fesc}

The escape of ionizing radiation from star-forming galactic disks is key to
understanding the cosmic reionization and intergalactic UV background.
However,
constraints on the galaxy-scale escape fraction remain highly uncertain
\citep{dayal18}. Here, we compare the fraction of ionizing photons that escape
the domain to the fraction that are being absorbed by gas (neutral hydrogen) and dust. The
instantaneous escape fraction of ionizing radiation is estimated as
$f_{\rm esc,i} = (Q_{\rm exit,i} + Q_{\rm lost,i})/Q_{\rm i}$, where
$Q_{\rm exit,i}$ is the rate of ionizing photons that exit through the vertical
boundary of the computational domain, and $Q_{\rm lost,i}$ is the ionizing
photon rate of ``lost'' photon packets that are terminated by the condition
$d_{\rm xy} > d_{\rm xy,\rm max}=1024 \pc$.\footnote{We have
  verified that $f_{\rm esc,i}$ converges to within $0.3\%$ when
  $d_{\rm xy,\rm max} \ge 1024 \pc$.} The gas and dust absorption
fractions are calculated as $f_{\rm gas,i} = \int \mathcal{I}_{\rm phot} dV/\Qi$
and $f_{\rm dust,i} = \int \mathcal{D}_{\rm i} dV/\Qi$, respectively, where
$\mathcal{D}_{\rm i} = \nH\sigma_{\rm d,i}c\mathcal{E}_{\rm i}/(h\nu_{\rm i})$
is the local dust absorption rate.

The solid orange line in row \textsf{(c)} of \autoref{f:hst} shows that the
instantaneous escape fraction varies with time significantly, with the
cumulative escape fraction ($\int Q_{\rm esc,i} dt /\int Q_{\rm i} dt$) of
$1.1\%$ (orange dashed line). The large temporal fluctuation in $f_{\rm esc,i}$
arises because (1) ionizing sources have short lifetimes, and the total $Q_i$ is
dominated by a small number of sources, (2) ionizing photon sources are near the
midplane, where dense gas absorbs most of the photons and blocks large volumes
of distant gas, and (3) complete escape of ionizing photons from the galaxy
requires very low-density pathways extending over several kpc, created by strong
hot winds powered by multiple recent supernovae \citep{dove2000}. Since the low
density channels must align favorably with young, unshielded ionizing sources,
conditions allowing ionized photon escape occur only rarely and intermittently.
While there is no significant correlation between $\Phi_{\rm i}$ and
$f_{\rm esc,i}$, we find that relatively high escape fraction ($\gtrsim 3\%$)
occurs only if $\Phi_{\rm i} > 5\times 10^{49}\,{\rm kpc^{-2}}\,{\rm s}^{-1}$. This
trend is expected given that higher $\Phi_{\rm i}$ allows ionizing photons to
penetrate larger distances, increasing the chances of escape
\citep[e.g.,][]{dove1994,dove2000,haffner2009}. Our time-averaged escape
fraction is in agreement with the observational estimate of $\sim 1$--$2\%$
based on \Halpha\ emission from high velocity clouds
\citep[e.g.,][]{bland-hawthorn2002}.

The escape fraction of non-ionizing radiation $f_{\rm esc,n}$ is limited only by
dust absorption and is significantly higher than $f_{\rm esc,i}$, with a
time-averaged (cumulative) escape fraction of $\sim 22\%$ (see Column (7) of
\autoref{tab:summarystats}). The large discrepancy for galaxy-scale escape fraction
between ionizing and non-ionizing radiation is in contrast to the results found
from radiation hydrodynamic simulations of star-forming molecular clouds. In
particular, \citet{kim2019} found that $f_{\rm esc,n}$ and $f_{\rm esc,i}$ are
quite similar over most of the evolution, for a range of cloud masses and sizes. In conditions of a star-forming cloud, the ionization parameter is 
much higher, which results in similar values of $f_{\rm esc,n}$ and $f_{\rm esc,i}$ (see below).

Row \textsf{(d)} of \autoref{f:hst} shows that dust grains contribute
significantly to the absorption of ionizing photons, reaching up to $\sim50\%$
of events globally. We find that the majority of absorption events
by both neutral hydrogen and dust occur in high density gas within $200\pc$ from
the midplane. Globally, 50\% of the absorption by neutral hydrogen occurs at
density above $\nH =15\cm^{-3}$, while 50\% of the absorption by dust occurs at
density above $\nH=56\cm^{-3}$.

%
\begin{figure}[t!]
  \center{\includegraphics[width=\linewidth]{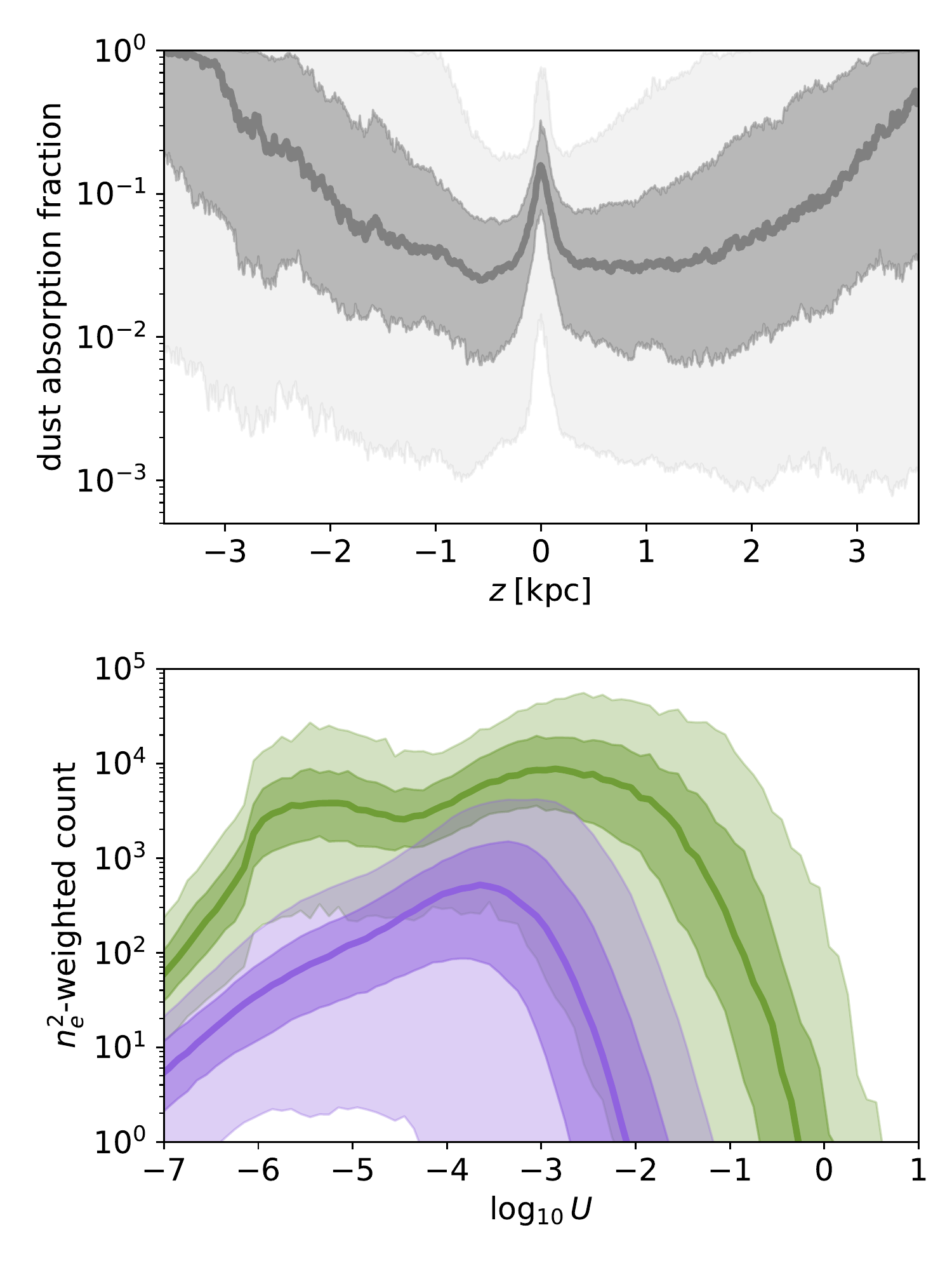}}
  \caption{\textit{Top:} Time-averaged, horizontally averaged fraction of
    ionizing photons that are absorbed by dust grains, $q_{\rm i, dust}$, as
    defined in \autoref{e:qi}, as a function of height from the midplane. The
    heavy curve shows the median profile, while the shaded regions bound the
    25$^{\rm th}$ to $75^{\rm th}$ percentiles (dark shaded region) and
    $5^{\rm th}$ to 95$^{\rm th}$ percentiles (light shaded region).
    \textit{Bottom:} Time-averaged distribution of the local ionization
    parameter $U = \mathcal{E}_{\rm i}/(h\nu_{\rm i}\nH)$, weighted by the local
    photoionization rate for gas within (green) and above (purple) 200\pc{} from
    the midplane. For both, the shaded regions bound the regions between the
    25\thh{} to 75\thh{} and 5\thh{} to 95\thh{} percentiles. The majority of
    photoionization (and recombination) takes place in dense gas near the
    midplane, which has a systematically higher ionization parameter than gas at
    high altitudes. This creates the overall shape seen in the top panel, where
    the relative importance of dust is highest near the midplane.}
\label{dustabs}
\end{figure}

\autoref{dustabs} characterizes the relative importance of dust absorption vs.
photoionization. In the top panel, we show the time-averaged (median) vertical
profile of the area-averaged dust absorption fraction, defined as
\begin{equation}\label{e:qi}
  q_{\rm dust,i} = \frac{\langle \mathcal{D}_{\rm i} \rangle}
  {\langle \mathcal{I}_{\rm phot} \rangle + \langle \mathcal{D}_{\rm i}
    \rangle}\,.
\end{equation}
Most of the time the dust absorption rate is a factor of $5$--$10$ lower than the
photoionization rate near the midplane, and a factor of $\sim 50$ lower at
$|z| \gtrsim 1\kpc$. This low median value of $\langle q_{\rm dust,i}\rangle $
might seem hard to reconcile with the time-averaged (cumulative) global dust
absorption fraction of 57\% (Column (6) of \autoref{tab:summarystats}). However, we
note that the majority of dust absorption events take place in dense gas (of
size a few tens of pc) near bright sources, whose vertical position changes from
snapshot to snapshot. As a result, the distribution of
$q_{\rm dust,i}$ at each height is strongly skewed toward high
values; 
the median value represents the typical absorption fraction in
``diffuse'' part of the ISM.

Assuming that WIM gas is near-fully ionized ($\nelec \approx \nH$) and in
photoionization--recombination equilibrium
($\mathcal{I}_{\rm phot} \approx \mathcal{R}$), one can show that the local dust
absorption rate is greater than the photoionization rate if
\begin{equation}
  U \gtrsim \alphaB/(c\sigma_{\rm d,i}) \sim 10^{-2}
\end{equation}
where $U \equiv \mathcal{E}_{\rm i}/(h\nu_{\rm i} \nH)$ is the the local
ionization parameter \citep[e.g.,][]{dopita2003,kim2019}. The bottom panel of
\autoref{dustabs} shows the time-averaged distribution of $U$ weighted by the
local recombination rate, for warm gas within (green) and above (purple) \hsep{}
from the midplane. Most of \Halpha\ emitting ionized gas is at low altitudes (as
shown by the distribution of $H_{\nesq{}}$ in \autoref{f:hst}, row
\textsf{(e)}), where the ionization parameter is $U \sim 10^{-4}$--$10^{-1.5}$.
In contrast, the WIM at high altitudes has a systematically lower ionization
parameter $U \sim 10^{-5}$--$10^{-3}$.\footnote{The bump at $U \sim 10^{-6}$
  comes from the partially ionized gas ($\xn > 0.1$) at warm--hot interfaces
  ($T \sim 2\times 10^4\Kel$), where most of ionizing photons are absorbed by
  gas if an ionizing source resides in a hot bubble.} These differences in
ionization parameter explain the relative roles of dust and gas in absorbing
ionizing photons at the midplane (where $U$ is larger and dust absorption can
exceed gas absorption) vs. high altitudes (where $U$ is smaller and absorption
by gas always dominates).

\subsection{Vertical Profiles and Volume Filling Factors}\label{s:vstructure}

\begin{figure*}[t!]
  \center{\includegraphics[width=\linewidth]{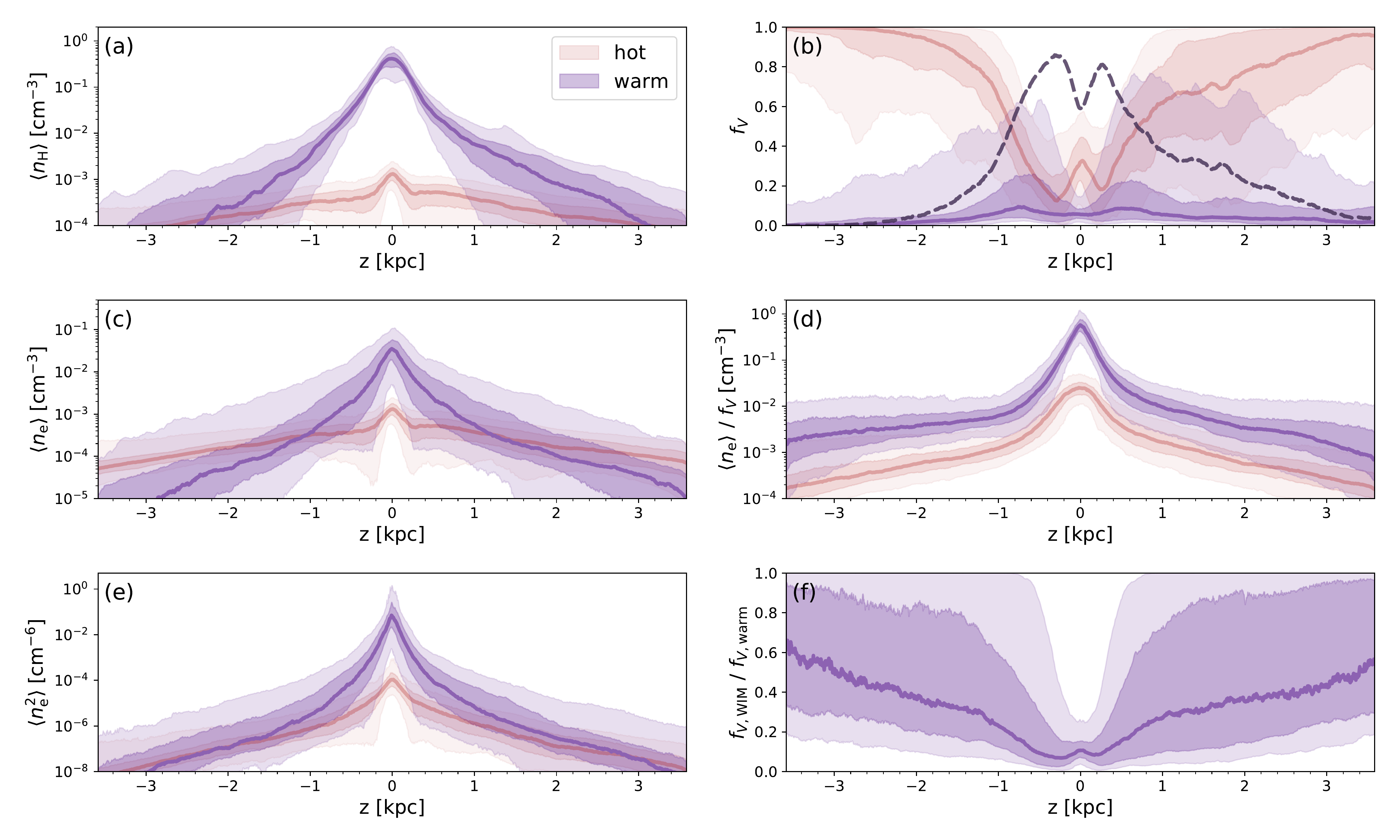}}
  \caption{Time-averaged, horizontally-averaged $z$ profiles of $\nH$ (panel \textsf{(a)}),
    the volume filling factors $f_V$ (panel \textsf{(b)}), $\nelec$ (panel \textsf{(c)}), the
    characteristic electron density ($\langle \nelec{} \rangle / f_{V}$, panel
    \textsf{(d)}), $\nesq$ (panel \textsf{(e)}), and the volume fraction of ionized gas within the warm medium 
    ($f_{V, \rm WIM} / f_{V, \rm warm}$, panel \textsf{(f)}). In each panel, the purple
    curve shows the median profile of the (selected) warm gas
    ($5050 \Kel < T < 2\times 10^4\Kel$), while the red curve shows the median
    profile of the hot gas ($T > 2\times 10^4 \Kel$). Dark and light shaded
    regions indicate the 25\thh{}--75\thh{} and 5\thh{}--95\thh{} percentiles.
    In \textsf{(b)}, the solid curve shows the volume filling factor for the WIM, while the dashed curve shows the volume filling factor for
    all warm gas, regardless of the ionization state.}\label{f:mzprofiles}
\end{figure*}

\begin{figure*}[t!]
  \center{\includegraphics[width=\linewidth]{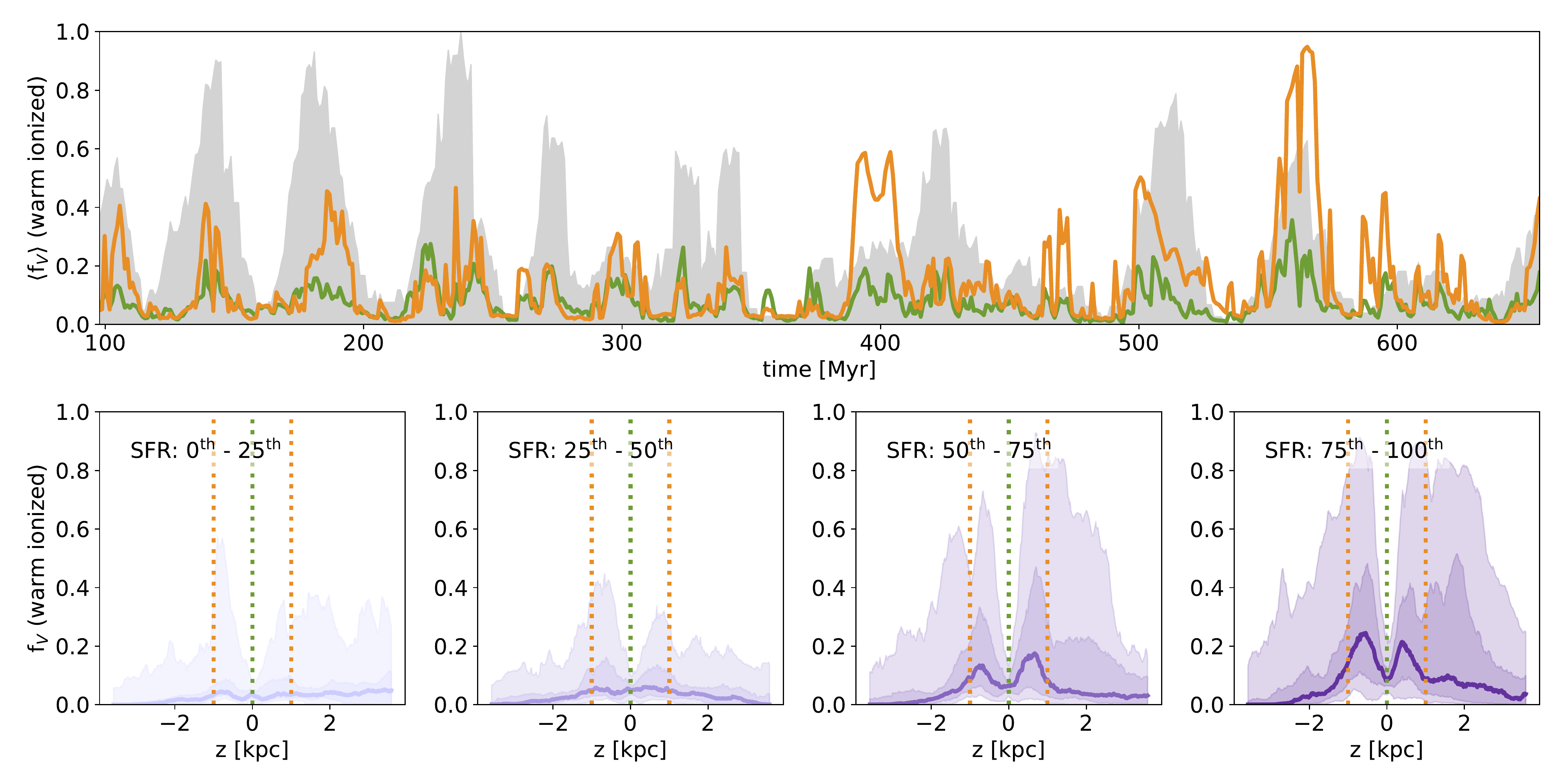}}
  \caption{\textit{Top:} Horizontally averaged volume filling fraction of warm
    ionized gas ($f_{V,{\rm WIM}}$) as a function of time for gas at
    $|z|<200 \pc$ (green) and for gas at $800\pc<|z|<1200\pc$ (orange). The grey
    shade is the rescaled SFR for comparison. \textit{Bottom:} the average
    z-profile of $f_{V,{\rm WIM}}$ in bins of $\Sigma_{{\rm SFR},{\rm 10 Myr}}$.
    The solid curve shows the median z-profile, while the shaded regions show
    the 25\thh{}-75\thh{} and 5\thh{}-95\thh{} percentiles. The dotted lines
    show the central positions of the slices shown in the top panel.}
\label{fvsfr}
\end{figure*}

As described in \citet{kim2018,vijayan2019}, spatio-temporally correlated SNe in
our simulation launch multiphase outflows consisting of hot winds and warm
fountains. Although hot winds attain high enough velocity ($ >200 \kms$ at
$|z| > 1\kpc$) to develop into galaxy-scale winds, the velocity distribution of
warm outflows is exponential with the typical outflow velocity of $\sim 60 \kms$
at $|z| = 1 \kpc$. This is insufficient to escape from the gravitational
potential well of the Milky Way, and as a result, most of warm outflows
eventually fall back toward the midplane as inflows. The vertically stratified density
profile results from the weight of gas balancing
the Reynolds stress associated with the outflow momentum flux (plus thermal and
magnetic pressure support).

\autoref{f:mzprofiles} shows as solid lines the time-averaged (median) vertical
profiles of $\langle \nH \rangle$, $\langle\nelec \rangle$,
$\langle n_e^2\rangle$, the volume 
filling-factor\footnote{For example, the volume filling-factor of warm ionized gas is
  defined as $f_{V,{\rm WIM}} = \int \Thetaw x_{\rm i} dA/\int \Thetaw dA$ where
  $\Thetaw{}$ is a top hat function that selects warm gas with
  $5\times 10^3 \Kel < T < 2\times 10^4 \Kel$. 
  Note that $f_{V, \rm WIM}$ is not 
  equivalent to the commonly used
  observational line-of-sight averaged
  filling factor derived from EM and DM toward pulsars under the assumption of constant electron density in ionized clouds
  \citep[e.g.,][]{reynolds1991a, berkhuijsen2006}.}  $f_{V}$, $\langle \nelec\rangle/f_{V}$, and $f_{\rm V,WIM}/f_{\rm V,w}$, where
$\langle \rangle$ refers to the area-average over the $x$-$y$
plane. The warm gas profiles are
shown in purple, while hot gas profiles are shown in red. Note that
$\langle \nelec\rangle/f_V$ is the density of ionized gas averaged over the
volume occupied by itself, i.e., it is the characteristic local density of
ionized gas.

The time-averaged (median) midplane densities of warm and warm ionized gas
(panels (a) and (c) in \autoref{f:mzprofiles}) are at $0.41 \pcc$ and
$0.032 \pcc$, respectively, as measured within $50\pc$ of the midplane. In panel
(b), the volume filling factors of total warm gas (dashed line) and WIM (purple)
show depressions near the midplane as this is where most hot gas is generated
via shock heating by supernovae. The total warm gas volume filling factor peaks
at $f_{V,{\rm warm}} \sim 0.86$ near $|z| \sim 300 \pc$. At $z \gtrsim 2 \kpc$,
the volume filling factors of both warm and warm ionized gas
(\autoref{f:mzprofiles}(b)) become increasingly small as the box becomes
dominated by the hot winds. However, the share of warm gas that is ionized (i.e.
$f_{V,{\rm WIM}}/f_{V,{\rm w}}$) increases as a function of distance from the
midplane (\autoref{f:mzprofiles}(f)). 
The characteristic number density
$\langle \nelec\rangle/f_V$ for WIM ranges between $9\times10^{-3} \pcc{}$
(25\thh{} percentile) and $5\times10^{-2} \pcc$ (75\thh{} percentile) for
$0.2 \kpc < |z| < 1 \kpc$ (\autoref{f:mzprofiles}(\textbf{d})). The vertical
profile of $\langle \nelec^2 \rangle$ (\autoref{f:mzprofiles}(\textbf{e})) for
warm gas is sharply peaked around the midplane, suggesting that most of \Halpha\
emission from warm ionized gas would originate near the disk midplane.

The volume filling factor of warm ionized gas is also correlated, albeit with
large temporal variance, with the global SFR in the box. The top panel in
\autoref{fvsfr} shows the time evolution of the average volume filling factor of
warm ionized gas within $200 \pc$ from the midplane (green) and within $200\pc$
from $z=1 \kpc$ (orange). For reference, the grey shaded area shows
$\Sigma_{\rm SFR,10\Myr}$ (scaled such that $\max(\Sigma_{\rm SFR,10\Myr}) =1$).
In general, the warm ionized gas volume filling factor is relatively small near
the midplane, with a median [25\thh{}, 75\thh{}] value of 0.057 [0.031, 0.097] at
$|z|<200$pc.

In contrast to the midplane region, the volume filling factor of the WIM at
$0.8\kpc < |z| < 1.2\kpc$ exhibits relatively large temporal fluctuations. In
particular, the warm ionized gas near $1 \kpc$ off the midplane accounts for the
majority of the volume ($f_{V,{\rm WIM}}>0.5$) for 8.4\% of the timesteps, and
accounts for at least 25\% of the volume ($f_{V,{\rm WIM}}>0.25$) for 18\% of
the timesteps. However, we note that the majority of our snapshots are not
dominated by WIM at $z=1 \kpc$; the median (mean) $f_{V,{\rm WIM}}$ at this
height is 0.062 (0.15).

In the bottom panels of \autoref{fvsfr} we show the average distribution of
$f_{V,{\rm WIM}}$ as a function of height off the midplane, binned by quartile
of the SFR ($10 \Myr$ average). Although there is not a strict correspondence
between the SFR and the volume filling factor of the warm ionized gas near
$1 \kpc$, the volume filling factor tends to rise with increasing SFR, in
particular the high-$f_V$ tail of the distribution.

\subsection{Scale heights of ionized gas emission}\label{s:scaleheight}

%
%
%

\begin{figure*}[t!]
  \center{\includegraphics[width=\linewidth]{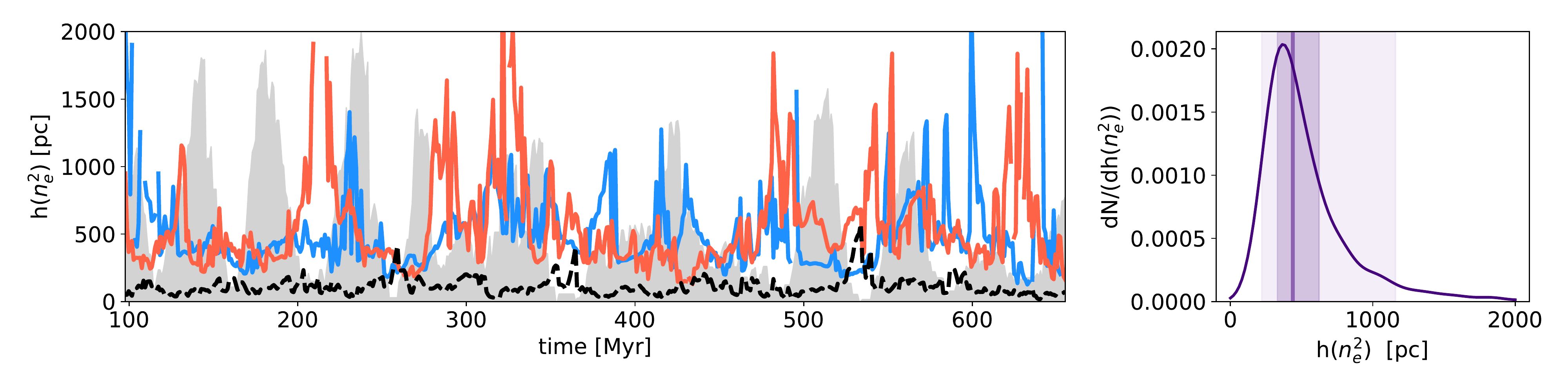}}
  \caption{\textit{Left:} The time evolution of the exponential-fit scale height
    $h(\nesq)$. The exponential-fit scale height includes only gas at $|z|>1$ kpc. The red and blue curves show the scale
    height above and below the midplane, respectively. The black dashed curve
    shows the $\nesq{}$ rms scale height ($H_{\nesq{}}$), defined as in \autoref{e:scaleheight}. The grey
    shade shows the (scaled) SFR for comparison. \textit{Right:} The
    distribution of the exponential-fit scale height over all time, both above and
    below the midplane. The thick purple line shows the median value of the scale
    height, while the shaded regions show the 25\thh{} to 75\thh{} percentile range.}\label{f:hnesq}
\end{figure*}

In this study, the scale height of warm ionized gas is measured in two different
ways: (1) the rms distance from the midplane ($H_{\rm w,i}$ and $H_{\nesq}$, as described in \autoref{s:H});
and (2) a fit of the vertical profile of $\nesq$ to an exponential 
function above some height from the midplane. In
this section, we describe our procedure and results for the latter, which is more relevant to 
existing measurements of the
DIG in external galaxies.

Most extragalactic observational studies measuring scale heights of the DIG from
\Halpha\ emission exclude the region closest to the midplane due to concerns
regarding contamination from \HII\ regions, dust extinction, and beam smearing
\citep[e.g.,][]{levy2019,boettcher2019}. To make a more fair comparison to
extragalactic measures of the \Halpha\ scale height, we fit an exponential
profile to the vertical profile of $\langle \nesq \rangle$, considering only the
high-altitude region with $|z| > 1 \kpc$. The left panel of \autoref{f:hnesq}
shows the time evolution of this exponential-fit $\nesq$ scale height,
$h(\nesq)$, measured for regions above (red) and below (blue) the midplane. In
the right panel, we show the probability density function of $h(\nesq)$
marginalized over time and direction, obtained from kernel density estimation
with a bandwidth of 0.25 (following Scott's Rule, \citealt{scott2015}). The
$\nesq{}$ exponential-fit scale height exhibits large temporal fluctuations in
the range $\sim 0.2$--$2.0 \kpc$, with no apparent correlation with recent star
formation activity (grey shades). The distribution of $h(\nesq)$ is right-skewed
with a median value $437 \pc$.

Because we have aggressively masked regions of box close to the midplane when
computing this observational scale height, the overall trend effected by the
exponential fit method is to increase the measured scale height (note that the
median rms scale height of $\nesq$ is only $94\pc$). Interestingly, the tail of
high exponential-fit scale heights ($h(\nesq)>1 \kpc$) is not correlated with
large $H_{\nesq}$ (as defined by \autoref{e:scaleheight}). We note, as a
caution, that there are significant differences in the \Halpha\ scale height as
measured by an external observer ($h(\nesq)$) and the scale height of WIM as
defined by \autoref{e:scaleheight} for the same snapshot. As shown in
\autoref{f:hnesq}, because the exponential-fit scale heights do not include the
inner, steeper regions of the $\nesq{}$ profile (\autoref{f:mzprofiles}c), they
are systematically larger than the time-equivalent rms scale heights.

\subsection{Distribution of $\nelec$}\label{s:nedist}

\begin{figure*}[t!]
\center{\includegraphics[width=\linewidth]{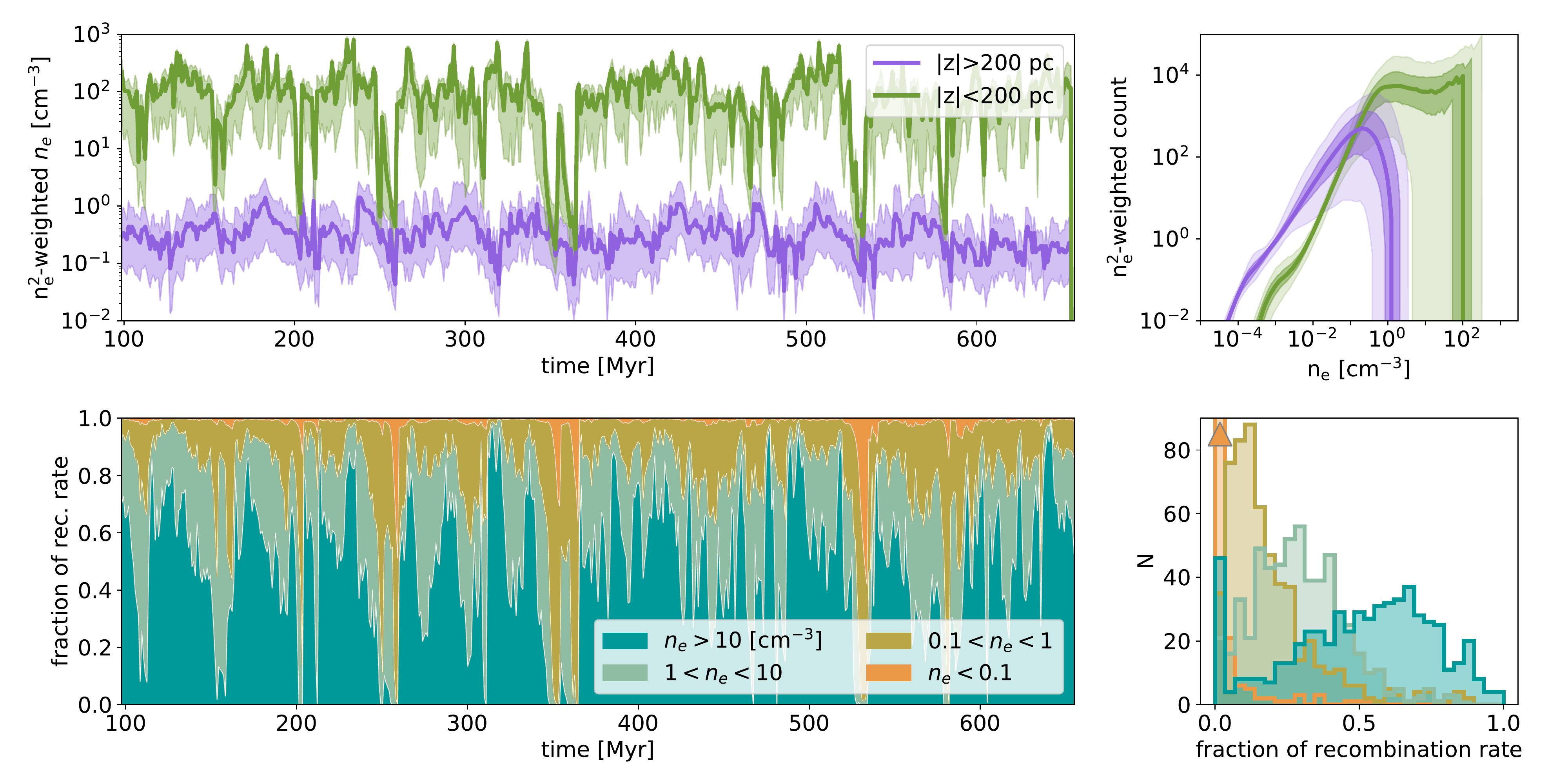}}
\caption{\textit{Top left:} time evolution of the electron density weighted by
  $\nesq$ for gas at $|z|<200\pc{}$ (green) and $|z|>200\pc{}$ (purple). The
  dark curve shows the median value of the $\nelec$ distribution, while the
  shaded regions show the 25\thh{} and 75\thh{} percentiles. \textit{Top right:}
  the time-averaged median distribution of recombination rate-weighted $\nelec$
  distribution. The shaded regions show the 25\thh{} to 75\thh{} percentile and
  5\thh{} to 95\thh{} percentile regions. \textit{Bottom left:} time evolution
  of the fraction of the \Halpha\ recombination rate contributed by gas of a
  given electron density slice (see legend). \textit{Bottom right:} the
  distribution over time of the recombination rate fractions shown at left.
  Again, the colors indicate slices in $\nelec$. The lowest density bin
  contributes $<3.4\%$ of the total recombination rate for $>90\%$ of the
  timesteps (N=523). For visual clarity, the figure is truncated at N=90; the
  orange triangle indicates that the lowest density bin extends to
  N=523.}\label{f:nedist}
\end{figure*}

To explore what fraction of \Halpha\ emission originates from low- versus
high-density gas, we calculate the density distribution of warm ionized gas
weighted by $\nesq$, which we take as a proxy for the local \Halpha\ emission
rate. In \autoref{f:nedist}, the top-left panel shows the time evolution of the
median and 25\thh{} and 75\thh{} percentiles in the distributions at height
$ |z| < 200 \pc$ (green) and at $|z| > 200 \pc$ (purple); the bottom-left panel
shows the fraction of total recombination that originates from gas at different
density slices. Right panels show the distributions over all time of $n_e$ (top)
and the contribution to the recombination rate (bottom).

As expected, the total emission is dominated by low-altitude gas. Although the
contribution of relatively dense gas ($\nelec > 10 \pcc$) dominates the total
recombination rate budget, the moderate-density ionized gas with
$1\pcc < \nelec <10 \pcc$ also contributes significantly to the total. At
$|z|<200\, \pc$, each logarithmic density interval above $\nelec \gtrsim 1 \pcc$
contributes approximately equally to the \Halpha\ emission. At $|z| > 200\pc$,
most of \Halpha\ emission comes from gas with $0.1\pcc <\nelec < 1 \pcc$, but it
accounts for, on average, only $\sim\! 14 \%$ of the total emission. The typical
density of \Halpha\ emitting gas at high $|z|$ is roughly consistent with the
observational estimate of the WIM density in the Solar neighborhood
\citep[e.g.,][]{berkhuijsen2008}.

\subsection{Distribution of EM}

\begin{figure}[t!]
\center{\includegraphics[width=\linewidth]{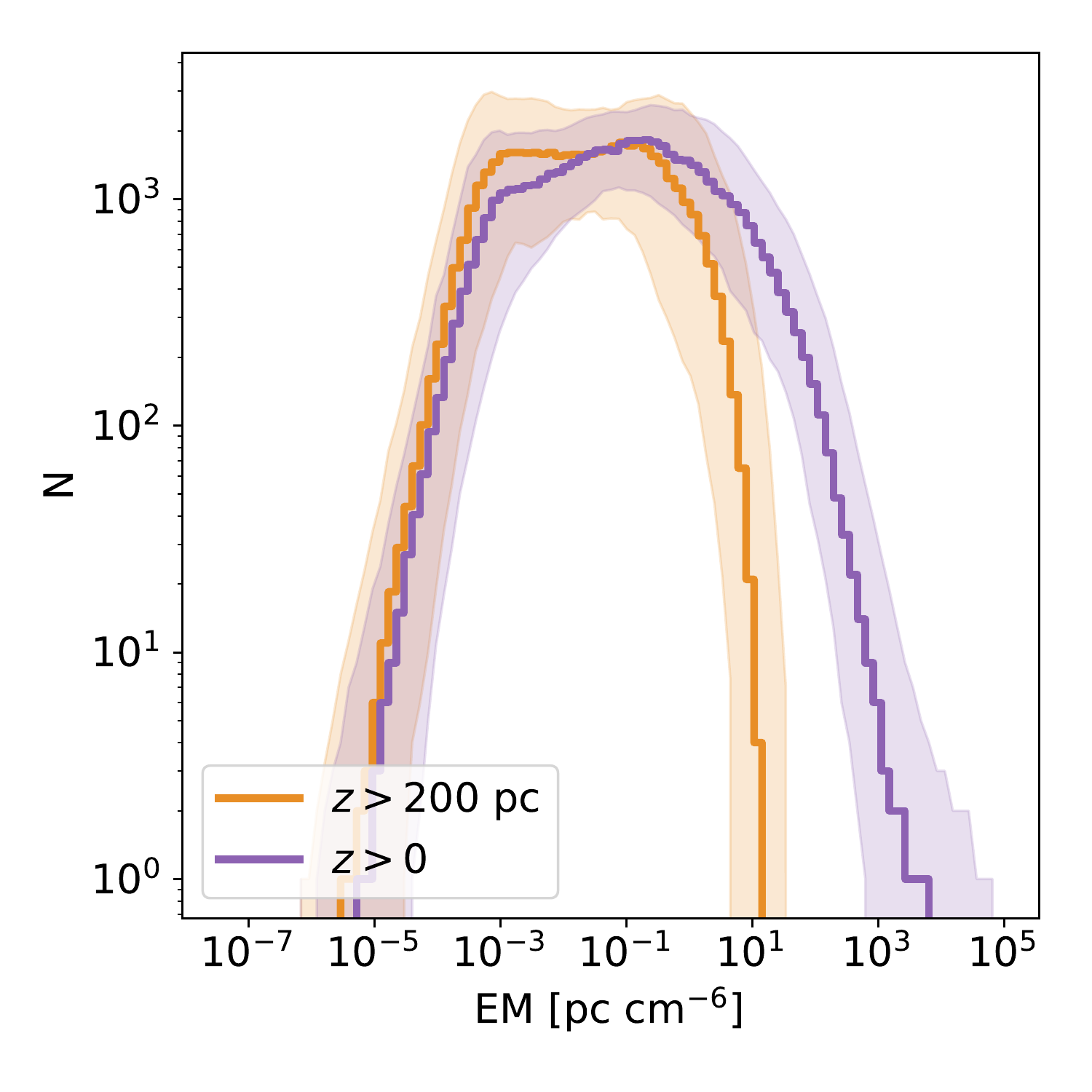}}
\caption{The distribution over time of EM ($=\int \nesq{} dz$) integrated over the plane of the disk,
  for an observer looking perpendicular to the plane of the disk and away from
  the midplane. The median distribution for an observer embedded in the midplane
  ($z>0$, purple) and an observer $200\pc$ away from the midplane ($z>200$ pc,
  orange) are shown by the thick stepped curves. The shaded regions show the
  25\thh{} and 75\thh{} percentiles for both cases. The high-EM tail of the
  midplane observer (purple) is indicative of dense, high EM regions near the
  midplane; these are analogous to classical \HII\ regions.}\label{f:emhst}
\vspace{20pt}
\end{figure}

In \autoref{f:emhst}, we show the time-averaged (median) EM distribution
integrated outward along the $z$-axis from $z=0$ (purple) and $z=200\pc$
(orange). In both cases, we take the (square) beam size to be the same as the
grid resolution $\Delta x = 4 \pc$. The two distributions are similar at low EM
($\lesssim 1 \,{\rm pc}\,{\rm cm}^{-6}$). At high EM, however, the two
distributions sharply diverge, with the observer at the midplane seeing
significantly more high EM instances than its counterpart at $200 \pc$. This is
a result of the contribution of dense gas near the midplane (see also
\autoref{f:nedist}). The high-EM extension is equivalent to the contribution
from classical \HII\ regions, though we do not presently resolve such regions or
model them self-consistently with dynamics. Because of the sensitivity of the EM
distribution to the presence of dense material near the midplane, we note that
it is crucial to fully exclude gas near the midplane in order to properly sample
extraplanar warm ionized gas.

\begin{figure}
\center{\includegraphics[width=\linewidth]{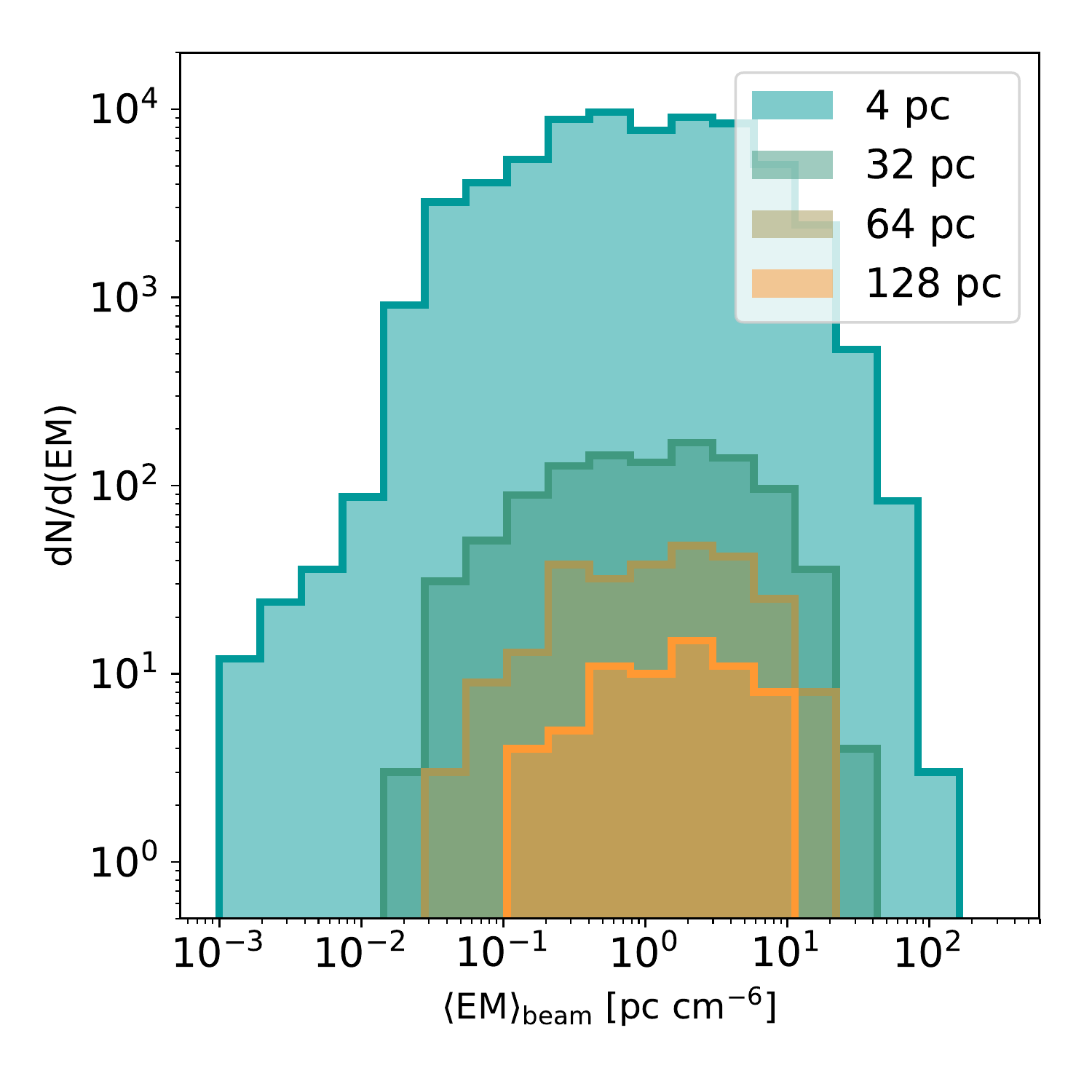}}
\caption{For a representative snapshot ($t=192 \Myr$), the effect of increasing
  the effective physical beam size from between the fiducial resolution
  ($4 \pc$) to $128 \pc$. As the beam size increases, the width of the EM
  distribution shrinks while the mean value remains unchanged. This effect
  complicates comparison to the observed width of the (WHAM) EM distribution, as
  there is no single physical scale that corresponds to the angular beam size in
  observation.}
\label{f:EMbeam}
\end{figure}

We now consider whether the width of the distribution of our EM measurements can
be used to gauge the agreement with our results and measurements taken of the EM
distribution of the WIM in the Milky Way. \citet{hill2008} found that the
distributions of the vertical component of EM from the WHAM survey is well
characterized by a lognormal distribution with mean
$\langle \log_{10}{\rm EM}_{\perp} ({\rm pc}\,{\rm cm}^{-6})^{-1}\rangle =
0.146$ and width $\sigma_{\log_{10}{\rm EM}_{\perp}} =0.19$. When considered
with an effective beam size of $4\pc$, the width of our EM distribution does not
match that of EM distributions for the solar neighborhood \citep{hill2008}.
However, this apparent width is degenerate with the physical size of the beam in
question. As shown in \autoref{f:EMbeam}, increasing the number of cells that
are considered in the measurement of the EM for a given line of sight decreases
the width of the resulting EM distribution, as the beam averages over a larger
area \citep[see also][]{berkhuijsen2015}. Because the beam size of the WHAM
observation is an on-sky beam size rather than physical beam size, we cannot use
the width of the EM distribution to test consistency of our models with the
Milky Way.

\subsection{The \Halpha\ line profile}\label{s:Haprofile}

\begin{figure*}[t!]
\center{\includegraphics[width=\linewidth]{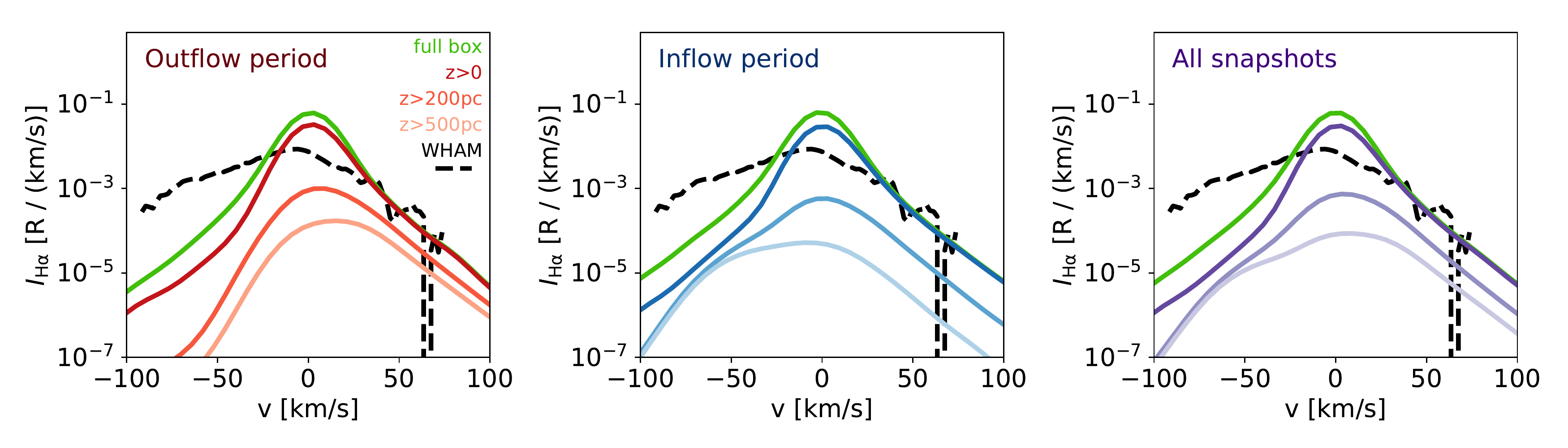}}
\caption{Time-averaged, horizontally averaged mean synthetic \Halpha\ line profiles for
  snapshots with bulk outflows at $z=1$\kpc{} (left, red), snapshots with bulk inflows
  at $z=1$\kpc{} (middle, blue), and all snapshots (right, purple). In each panel, the
  average WHAM line profile at $85\degr <b<90\degr $ is shown by the dashed
  black curve. All profiles are given as specific photon intensities in
  Rayleigh per $\kms$
  (${\rm R} = 10^6/(4\pi)\, {\rm photons}\,{\rm cm}^{-2}\second^{-1}{\rm
    sr}^{-1}$). The green curve shows the line profile as integrated across the
  full TIGRESS box; note that the presence of an outflow or inflow is determined
  by the top half of the box ($z>0$), and therefore does not necessarily
  correspond to an outflow or inflow in the bottom half of the box. The colored
  curves show, in order of increasing lightness, \Halpha\ line profiles as
  integrated from $z=0$, $z=200$ pc, and $z=500 \pc{}$ to the top of the box.}
\label{velprofile}\vspace{20pt}
\end{figure*}

\begin{figure*}
\center{\includegraphics[width=\linewidth]{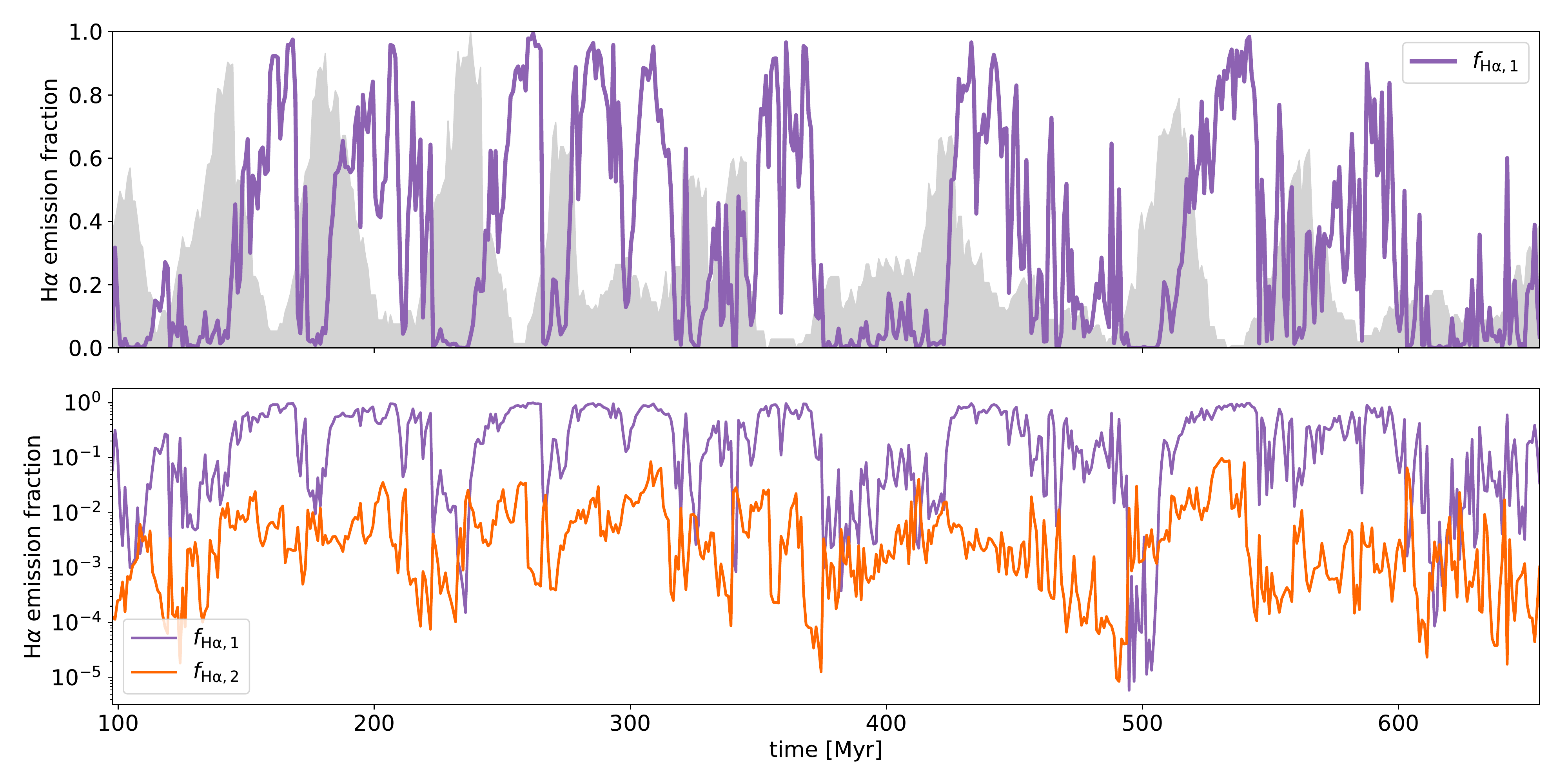}}
\caption{\textit{Top:} The fraction of the high velocity ($v_z > 50\kms$)
  \Halpha\ wing that is emitted from gas at $z>200\pc$ is shown in purple
  (\autoref{e:wing1}). For reference, the 10 Myr-averaged SFR is shown in grey.
  The apparent anticorrelation between the two suggests that the importance of
  high altitude \Halpha\ emission peaks approximately 20 Myr after the peak of
  star formation. \textit{Bottom:} the fraction of high velocity \Halpha\
  emission that originates from high altitudes is again shown in purple, as at
  top. The fraction of the total \Halpha\ emission that originates from high
  velocity gas is shown in orange (\autoref{e:wing1}).}
\label{highvelcontrib}\vspace{30pt}
\end{figure*}

Observations of high velocity gas have often been used to detect and
quantitatively characterize the properties of galactic outflows \citep[see,
e.g.][]{hill2008,wood2015,cicone2016,rodriguezdelpino2019}. The integrated
\Halpha\ line profiles constructed from TIGRESS thus both act as a benchmark for
the simulation as compared to \Halpha\ surveys of the Solar neighborhood, and
provide insight into the physical origin of high velocity gas seen in integrated
line profiles of external galaxies.

To construct line-of-sight integrated profiles, we first compute the \Halpha\
photon emissivity of each cell as
$j_{{\rm H}\alpha}(v) = (4\pi)^{-1} \alpha_{\rm eff,{\rm H}\alpha} \nelec^2
\phi(v)$, where
$\alpha_{\rm eff, H\alpha}=1.17\times10^{-13} T_4^{-0.942 - 0.030 \ln{T_4}}
\cm^3 \second^{-1}$ and the normalized line profile
$\phi(v)$ is a Gaussian with thermal width $\sim\!\! 9.1 \kms{}$ (for a pure
hydrogen gas at $T=10^4 \Kel$) centered at the vertical velocity $v_z$
\citep{draine2011a}. The line profile is obtained by integrating
$j_{{\rm H}\alpha}(v)$ along the line of sight perpendicular to the midplane
\begin{equation}
  I_{\rm H\alpha}(v) = \int_{z_{\rm min}}^{z_{\rm max}} j_{\rm H\alpha}(v) dz\,.
\end{equation}\label{eq:iha}
The effects of absorption and scattering by dust are ignored.

First, we consider mock ``observations'' made of the Milky Way DIG by
constructing \Halpha\ line profiles integrated from a given $z_{\rm min}$ to the
top of the box ($z_{\rm max}=+L_z/2$). For this exercise, we construct the line
profiles using the top half of the box only (i.e. an observer looking
perpendicular to the plane of the disk away from the midplane), and consider
$z_{\rm min}=0, 200\pc$, and $500\pc$ to separate contributions from dense and
diffuse ionized gas. We also include the profile as integrated across the full
box ($z_{\rm min}=-L_z/2$). In \autoref{velprofile}, we show the time-averaged
and horizontally-averaged mean line profiles for outflow states, inflow states
and for all snapshots. The line intensity is given in units of Rayleigh per
${\rm km}\,{\rm s}^{-1}$
(${\rm R} = 10^6/(4\pi)\, {\rm photons}\cm^{-2}\second^{-1}{\rm sr}^{-1}$). We
follow \citet{kim2018} in defining outflow states as snapshots in which
$\langle \dot\Sigma(1\kpc)\rangle > 10^{-3} \Msun\kpc^{-2}\yr^{-1}$, and inflow
states as snapshots in which
$\langle \dot{\Sigma}(1\kpc)\rangle < -10^{-3} \Msun\kpc^{-2}\yr^{-1}$, where
$\langle \dot\Sigma(1\kpc)\rangle $ is the area-averaged mass flux through the
$x$-$y$ plane at $z=1 \kpc$. In all panels, the mean WHAM \Halpha\ profile at
$85 \degr < b <90 \degr$ is shown by dashed black curves.

Though there is not a clear equivalent to the volume probed by WHAM, the peak
line intensity $I_{\rm H\alpha} \sim 10^{-2}{\rm R}/({\rm km}\,{\rm s}^{-1})$
seen in WHAM falls between the line profile integrated from the midplane and the
line profile integrated from $z_{\rm min} = 200\pc$. This indicates that the
high-intensity, low velocity component seen in the mock line profile is mostly
from dense ionized gas, which is excluded in the WHAM survey.

For both inflow-dominated and outflow-dominated periods, the positive-velocity
wing at $v>20\kms$ in the synthetic line profile is quite similar to that from
WHAM, indicating that the velocity distribution of outflowing gas in our
simulations is consistent with the local Milky Way. The high-velocity wing of
the \Halpha\ profile has an exponential shape, consistent with the exponential
mass distribution previously identified by \citet{kim2018} for high-altitude,
high-velocity gas in TIGRESS. Notably, much of the high-velocity wing originates
at high $z$ during outflow periods, but this is not the case during inflow
periods.

The overall simulated line shapes that most resemble the mean WHAM profile are
from the inflow period at $z>200\pc$. However, even for this period, the
observed average \Halpha\ profile is systematically wider than the time-averaged
simulated line profiles at negative velocities. The deficit in negative-velocity
emission in simulations compared to the WHAM profile offers intriguing support
for the idea that the observed ionized gas with large negative velocities
($v \lesssim - 50 \kms$) has extragalactic origin, which is not incorporated in
our simulation.

\subsubsection{The physical origin of high velocity gas}\label{sec:Hawing}

It is of observational interest to study (1) 
the possible link between the \Halpha\ emission from high-altitude outflowing gas and
recent star formation activity and 
(2) how well the high-velocity wing of the \Halpha\ line profile traces such outflowing gas.
To address these questions, we compute the fractional contribution of material at $z > 200\pc$ to
the high velocity wing emission
\begin{equation}
f_{\rm H\alpha,1} = \dfrac{\int^{\infty}_{v_{\rm min}} I_{{\rm H}\alpha}
    (v; z_{\rm min}=200\pc)dv}{\int^{\infty}_{v_{\rm min}} I_{{\rm H}\alpha}(v; z_{\rm min}=0)dv} \label{e:wing1}
\end{equation}
and the contribution of the wing to the total emission
\begin{equation}
f_{\rm H\alpha,2} = \dfrac{\int^{\infty}_{v_{\rm min}} 
I_{{\rm H}\alpha}(v; z_{\rm min}=0)dv}{\int^{\infty}_{-\infty} I_{{\rm H}\alpha}(v; z_{\rm min}=0)dv} \label{e:wing2}
\end{equation}
where we take $v_{\rm min} = 50\kms$, following \citet{kim2018}.

The top panel of \autoref{highvelcontrib} shows the time evolution of
$f_{\rm H\alpha,1}$, which suggests that peak of $f_{\rm H\alpha,1}$ lags the
SFR averaged over $10 \Myr$ (gray shades). This is expected if star formation
drives high velocity gas from the midplane, as there is a several-Myr delay
between star formation and supernova activity, and since a travel time of at
least $\sim 10 \Myr$ is needed for gas to escape the midplane regions of the
disk.

The bottom panel of \autoref{highvelcontrib} shows both $f_{\rm H\alpha,1}$ and
$f_{\rm H\alpha,2}$ in purple and orange, respectively. While the high-velocity
wing is always a small fraction of the total \Halpha, we find that the presence
of a wing that accounts for 2\% of the total emission
($f_{\rm H\alpha,2} > 0.02$) indicates that more than 50\% of the high-velocity
gas is likely to be extraplanar ($f_{\rm H\alpha,1} > 0.5$). However, the
converse is not true, and strong wing emission occurs in only a small number of
snapshots. Still, it is notable from the history of $f_{\rm H\alpha,1}$ shown in
\autoref{highvelcontrib} that at most times, more than half of the high-velocity
emission originates in the extraplanar region.

\section{Discussion}\label{s:discussion}

\subsection{Comparison with other numerical models}

Similar to our work, several studies
\citep{wood2010,barnes2014,barnes2015,vandenbroucke2018} investigated the
formation of DIG by post-processing the density grids taken from (M)HD
simulations of supernova-driven turbulent, multiphase ISM in a
vertically-stratified box \citep[with simulation inputs
from][]{joung2006,joung2009,hill2012,girichidis2016}. These studies have shown
that turbulence and superbubbles naturally produce low-density channels through
which ionizing photons can travel large distances and photoionize an extended
layer of warm neutral gas at high altitudes, which itself is produced by
supernovae-driven outflows. However, among these and our own study there are
important differences in modeling stellar feedback (in the MHD simulation) and
ionizing source properties (in the post-processing), which may lead to
consequential differences in density structure and the WIM distribution.

The simulations by \citet{joung2006,joung2009,hill2012} incorporated both
distributed (Type Ia and ``field'' Type II) and clustered supernovae (also
including early wind energy input). However, these simulations did not have
self-gravity and star formation was not directly modeled, so the rate of SNe was
imposed at a fixed value and the locations of single and clustered supernova
explosions were chosen randomly (horizontally), uncorrelated with gas density.
As a consequence, the SN explosions were not as effective as they should have
been in disrupting and blowing out dense structures in the midplane region. The
resulting vertical density structure in these MHD simulations was therefore more
centrally peaked around the midplane and had lower density at high altitudes
than observations suggest (e.g., Fig. 3 of \citealt{joung2006} and Fig. 1 of
\citealt{barnes2014}). These vertical structure discrepancies then affect
predictions for $n_e$ profiles and EM ($\propto n_e^2$) distributions (e.g.
Figs. 4, 6 of \citealt{wood2010}).

Recent controlled numerical experiments have shown that the details of supernova
feedback have a direct impact on the thermal phase balance, spatial distribution
and relative volume filling factors of gas phases in the disk, and launching of
outflows \citep[e.g.,][]{WalchSILCC2015,li2017,hill2018}. For example, the
outflow properties of warm fountains sensitively depends on the vertical scale
height of SNe (relative to the gas scale height), as the fraction of SNe that
interact with dense gas varies with the SNe scale height (e.g. \citealt{li2017};
see also appendix of \citealt{kim2018}). The volume filling factor of hot gas
vs. warm gas is also quite sensitive to the correlations of supernovae relative
to the gas density \citep{WalchSILCC2015}. In addition, the mass and volume
fractions of warm gas varies with the input FUV heating rate \citep{hill2018},
but the previous (M)HD simulations adopted a temporally constant FUV heating
rate (as well as SN rate).

In contrast to simulations previously used for modeling the DIG, the vertical
density distribution in our simulation is in much better agreement with
observations (see  \autoref{s:obs}), presumably because the self-gravity and
self-consistent treatment of star formation and SN+FUV feedback in TIGRESS leads
to a more realistic space-time correlation between gas density and the stellar
energy sources responsible for the thermal, turbulent, and magnetic pressure in
the ISM \citep{kim2017,kim2018}.

The Monte-Carlo photoionization post-processing simulations by
\citet{wood2010,barnes2014,barnes2015,vandenbroucke2018} set the number of
ionizing sources per area to $24 \kpc^{-2}$, to be consistent with observational
constraints \citep[e.g.,][]{garmany1982}; the positions of ionizing sources were
distributed randomly horizontally, but followed a Gaussian distribution with a
scale height of $63\pc$ in the vertical direction \citep{maizapellaniz2001}.
Rather than setting a photon input rate consistent with the adopted supernova
rate in the underlying (M)HD simulation, in these models the ionizing photon
rate per source ($Q_{\rm i,sp}$) was varied as a free parameter, ranging from
$\sim 10^{47}\second^{-1}$ to $10^{50}\second^{-1}$. The high end would
correspond to $\Sigma_{\rm SFR} \sim 2 \times 10^{-2} \Msun\pc^{-2}\Myr^{-1}$,
while lower values correspond to lower SFRs and/or a small fraction of photons
leaking from \HII\ regions.

In these studies, the input ionizing photon rate was shown to be the most
important factor determining the structure and extent of the WIM
\citep[e.g.,][]{wood2010,barnes2014}. While the moderate value
($Q_{\rm i,sp}\sim$ a few $\times 10^{49}\second^{-1}$) maintained both a
neutral disk and an extended DIG, a $Q_{\rm i,sp}$ that was too high (low)
resulted in an overabundance (underabundance) of ionized gas. Most of these
studies found that for realistic $Q_{\rm i,sp}$, the WIM density is lower and
\Halpha\ scale height is smaller than the observational constraints. The
exception is the model of \citet{vandenbroucke2018}, in which the extended DIG
is produced by cosmic ray feedback \citep{girichidis2016}. For
$Q_{\rm i,sp} = 4.26 \times 10^{49}\second^{-1}$, they found the exponential
scale height of the WIM $h(\nelec) \sim 1.4 \kpc$ and $h(\nesq) \sim 0.7\kpc$ at
$|z|> 500\pc$ and $\langle \nelec \rangle \sim 0.02\cm^{-3}$ at $|z|=1\kpc$,
which is in agreement with the observed Reynolds layer (see
\autoref{eq:Reynolds}).

In the post-processing radiation treatment adopted for the present study,
neither the locations nor the luminosities of ionizing radiation sources are set
arbitrarily. Instead, photon sources are the young cluster particles that form
as a result of self-gravitating collapse. The ionizing sources therefore have
realistic placement relative to the distribution of cold and warm clouds that
can absorb ionizing photons, and relative to the hot gas channels created by
supernovae that allow ionizing photons to travel long distances. The
luminosities of individual sources are set by the clusters' masses and ages.

%
%
%
%
%

\rev{Finally, it is of interest to compare our result to \citet{peters2017}, who
  conducted radiation hydrodynamic simulations of a star-forming galactic disk
  in which the dynamical effect of radiation feedback was self-consistently
  included by the adaptive ray tracing method.
  Compared to the TIGRESS simulation, their simulations 
  lack galactic shear
  and magnetic fields, but include complex thermochemistry coupled with radiative
  transfer. 
They also model the massive star population in each sink particle by directly
sampling from the IMF, which captures stochastic effects.
 It is important to note, however, that the simulation of
 \citet{peters2017} spans a total time of 70 Myr (and only 38 Myr after the
  first star formation), so it is not guaranteed that the simulation has reached a quasi-steady state.

\citet{peters2017} find that the inclusion of radiation feedback does not
  significantly affect the star formation rate surface density (as compared to their model with SNe and stellar winds). This conclusion is in line with  
  \citet{kannan2020}, who find that radiation pressure has a negligible impact on the 
  SFR surface density compared to a model with SNe and photoheating.
  They also find that after an initial transient, including photoheating (both non-ionizing and ionizing) has only a modest effect compared to a simulation with only SN feedback.  
  Overall, in solar neighborhood models, ionizing radiation feedback does not appear to be dynamically important.  This suggests that our results would not have been significantly altered if we had included time-dependent radiation feedback in the original TIGRESS solar-neighborhood simulation.   We remark, however, that in denser galactic environments than the solar neighborhood, ionizing radiation and other  ``early feedback'' might be more dynamically consequential, because more rapid dynamical contraction of clouds and efficient star formation could occur before the onset of SNe to  disperse gas.}

 \rev{
The results of  \citet{peters2017} on SFRs, ionizing photon production, and ionized gas content are similar to our own. 
Near the end of their simulation, 
they find $\Sigma_{\rm SFR} \sim 10^{-3} M_\odot$ yr$^{-1}$ kpc$^{-2}$,
within a factor of a few of the median values found in this work.
They find a median ionizing luminosity surface density of $\sim4\times10^{-4}$
erg s$^{-1}$ cm$^{-2}$, which somewhat smaller than our median ionizing luminosity
surface density ($1.2\times10^{-3}$ erg s$^{-1}$ cm$^{-2}$) and comparable to
our 25\thh{} percentile value ($5.5\times10^{-4}$ erg s$^{-1}$ cm$^{-2}$).
As in this work, they also find that the \Halpha\ emission is dominated by  recombinations in photoionized gas, 
  with significant temporal fluctuations on a timescale of a few Myr. 
  The mass fraction of ionized gas ($\sim 4$--$10\%$) is also in good
  agreement with our result.}

\rev{ 
\citet{peters2017} also report volume filling fractions
within 100 pc of the midplane for their simulations.
To make a comparison to their results, we recompute the
volume filling factor over the same temperature ranges
as they adopt, denoted by the subscript label ``${\rm -P}$'' (these temperature ranges different from our definitions).  
Below, superscript labels indicate
model name, where FRWSN is their run with radiation feedback included, and
FWSN is their run without radiation feedback.
We find a median [25\thh{}, 75\thh{} percentile] 
value of
$f_{\rm V,warm-P}$ of 0.55 [0.45, 0.64], in excellent agreement with
the simulation with radiation feedback included $f_{\rm V, warm-P}^{\rm FRWSN}=0.6$ 
at $t>50$ Myr. 
We find similarly good agreement for volume filling factors of other phases
that \citet{peters2017} considered
when radiation feedback is included.
In runs without radiation  feedback (i.e. only collisionally-ionized gas), \citet{peters2017}
found a much lower median volume 
filling factor, $f_{\rm V, warm-P}^{\rm FWSN} = 0.3$.
}

\subsection{Comparison with observations: the Dickey-Lockman and Reynolds Layers}\label{s:obs}

\begin{figure}[ht!]
  \center{\includegraphics[width=\linewidth]{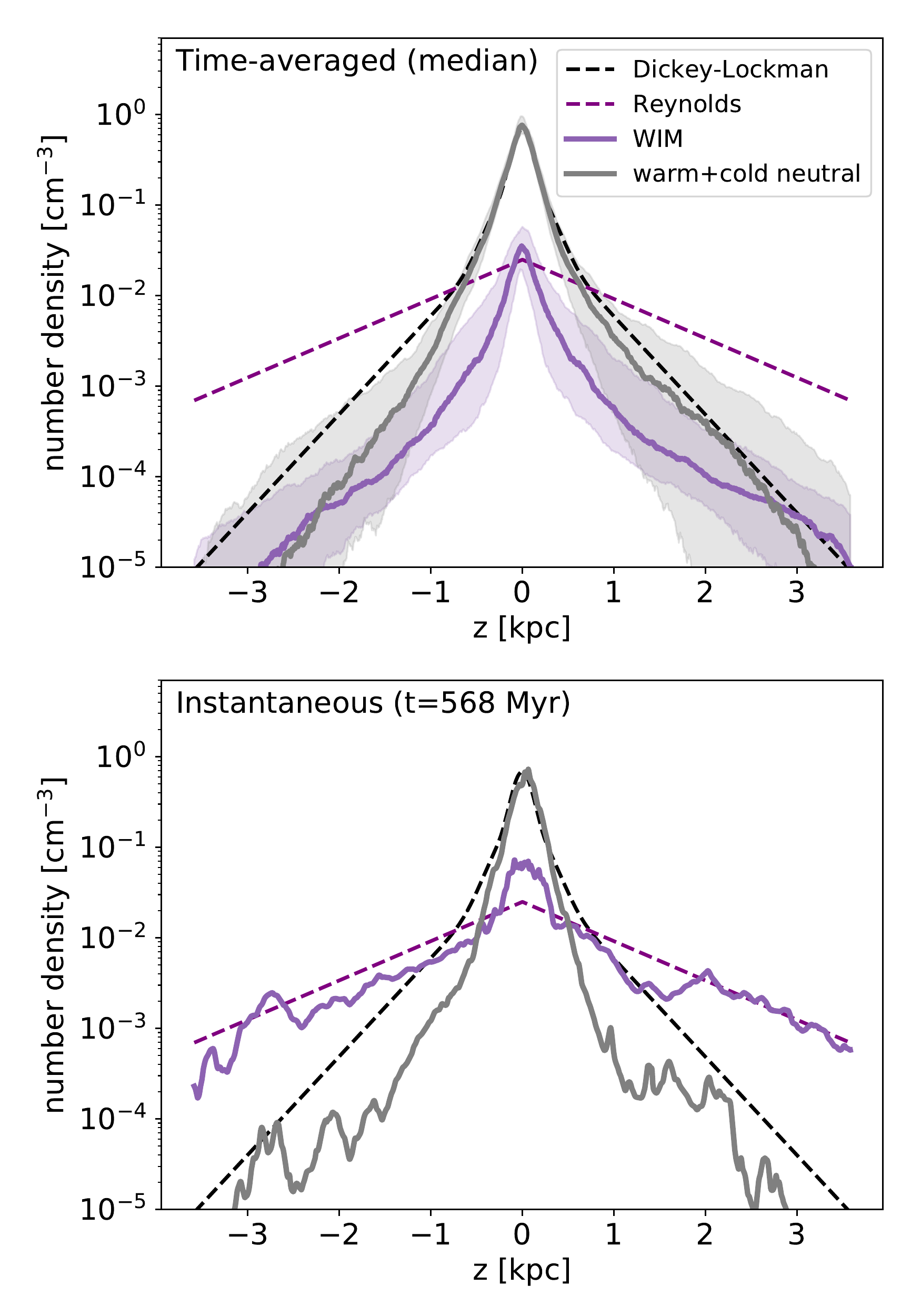}}
  \caption{Comparison of vertical profiles of $\nHI$ (grey) and $\nelec$
    (purple) from the simulation with Milky  Way  observations. The dashed lines show
    observed estimates of the Dickey-Lockman layer for the neutral hydrogen (see
    \autoref{eq:DL}) and the Reynolds layer for free electrons (see
    \autoref{eq:Reynolds}). The solid curves show the mean (top) and
    instantaneous (bottom, at $t=568\Myr$) TIGRESS $\nelec$ and $\nHI{}$
    vertical profiles. Shaded regions in the top panel indicate 25\thh and
    75\thh percentile range.}\label{f:zprofobs}
\end{figure}


Based on various surveys of $21 \cm$ emission from neutral atomic hydrogen,
\citet{mckee2015} estimated that the vertical distribution of \ion{H}{1} in the
solar neighborhood follows
\begin{align}
  \dfrac{\langle n_{\rm H^0} \rangle_{\rm D\mbox{-}L} }{{\rm cm}^{-3}}
  & = 0.47 e^{-\tfrac{1}{2}\left(\tfrac{z}{90\pc}\right)^2} + 
  0.13 e^{-\tfrac{1}{2}\left(\tfrac{z}{225\pc}\right)^2} \nonumber \\
  & \phantomrel{=} {} + 0.077
  e^{-\left(\tfrac{|z|}{403\pc}\right)}\,,\label{eq:DL}
\end{align}
where the two Gaussian components represent warm-cold \ion{H}{1} in the main
disk and the exponential component accounts for an extended layer at high
altitudes. This ``Dickey-Lockman'' profile\footnote{The functional form of
  \autoref{eq:DL} is suggested by \citet{dickey1990} to match \ion{H}{1}
  observations of the inner ($4\kpc \lesssim R_0 < 8.0 \kpc$) Galaxy.
  \citet{mckee2015} multiplied the densities of the Dickey-Lockman profile by
  1.2 to bring the total \ion{H}{1} column density to that of the solar
  neighborhood value $7.45 \times 10^{20}\cm^{-2}$ \citep{heiles1976}.} is shown
as black dashed lines in \autoref{f:zprofobs}.

The DM of pulsars with known distances provides a direct measure of the WIM
content in the Milky Way. The ratio between DM and the pulsar distance indicates
the line-of-sight average electron densities of $\sim 0.01$--$0.1\pcc$ for
pulsars with $|z| \lesssim 1\kpc$, with the vertical component of DM saturating
at ${\rm DM}_{\perp} \sim 25\pc\pcc$ for pulsars at $|z| > 1\kpc$
\citep[e.g.,][]{reynolds1991b,gaensler2008,schnitzeler2012,deller2019}. A number
of studies found that an exponential disk with (extrapolated) midplane density
$\langle n_{\rm e,0}\rangle \sim 0.01$--$0.03\pcc$ and scale height
$h(n_e) \sim 1\kpc$ reasonably accords with observations. The dashed purple
lines in \autoref{f:zprofobs} show the widely adopted form
\begin{equation}
  \langle \nelec \rangle_{\rm R} = 0.025 \exp(-|z|/1\kpc)
  \pcc \label{eq:Reynolds}
\end{equation}
for the Reynolds layer. Note that based on current observed estimates, the WIM
begins to dominate over \ion{H}{1} for $|z| \gtrsim 700\pc$.

The top panel of \autoref{f:zprofobs} shows that
that the time-averaged TIGRESS profile (solid grey curve) matches the observed
Dickey-Lockman profile for neutral gas quite well out to $|z|=3\kpc$ (i.e. the
observed profile lies within the 25\thh{}-75\thh{} percentile range of the
TIGRESS profile). As shown for example in the lower panel of
\autoref{f:zprofobs}, individual instantaneous snapshots also agree quite well
with the observed Dickey-Lockman profile within $|z|<500 \pc$ (median of the
logarithmic residual
$\log_{10}(\langle n_{\rm H^0}\rangle /\langle n_{\rm H^0}\rangle_{\rm
  D\mbox{-}L})$ is $0.11$). At larger $|z|$, there is more variation in time, as
indicated by the grey shaded region in the top panel.

In the top panel of \autoref{f:zprofobs}, the time-averaged median profile of
the WIM from our simulation is shown by the solid purple curve (the shaded
region again shows the 25\thh{}-75\thh{} percentile region). Clearly, the
normalization of our mean WIM profile at large $|z|$ is well below the
observational estimate of the Reynolds layer (dashed purple line), although the
slope of our WIM profile at heights $|z| \gtrsim 1\kpc$ becomes shallower than
that of the total warm gas, which is in agreement with observations.  In order
for the vertical profile of warm ionized gas to be significantly shallower than
that of the total warm gas, the ionization fraction of warm gas must rise as a
function of distance from the midplane -- indeed, this is shown in panel (e) of
\autoref{f:mzprofiles}.

Though the properties of the observed Reynolds layer are not reproduced by the
time-averaged WIM profile in our simulation, we note that profiles quite similar
to \autoref{eq:Reynolds} are recovered in a minority of snapshots. An example is
shown in the lower panel of \autoref{f:zprofobs}.

%
%

\subsubsection{Potential Explanations for Discrepancies with Observations}\label{s:discrepancies}

One can think of two possible reasons that can explain the discrepancy between
our median WIM profile and the observational estimate: (1) lack of ionizing
photons and (2) lack of high-altitude gas to be ionized. Regarding the first
possibility, massive stars in the TIGRESS simulation produce enough photons to
maintain the ionization of the Reynolds layer. For a clumpy WIM disk with an
exponential scale height $h$ and a constant volume filling fraction
$f_{V,{\rm WIM}}$, the minimum ionizing photon rate per unit area required to
balance the total recombination is
$\Phi_{\rm i,min} = \alphaB\langle n_{\rm e,0}\rangle^2 h/f_{V,{\rm WIM}}$,
where $\langle n_{\rm e,0} \rangle$ is the area-averaged electron number density
at $z=0$. Adopting $f_{V,{\rm WIM}} = 0.1$ \citep[e.g.,][]{berkhuijsen2008} and $\langle n_{\rm e,0} \rangle=0.025 \cm^{-3}$ (\autoref{eq:Reynolds}), the
ionization of the Reynolds layer requires
$\Phi_{\rm i, min, R} = 4.8 \times 10^{49}\second^{-1}\kpc^{-2}$.
This adopted value of $f_{V,\rm WIM}$ yields (obtained by integrating \autoref{eq:Reynolds}) ${\rm EM} = n_{\rm e,0}^2 h / (2f_{V, \rm WIM}) = 3.12 \pc{} \cm^{-6}$.
This is within a factor of two of the measurement of \citet{hill2008}, who find an emission
measure of $1.406 \pm 0.004 \pc{} \cm^{-6}$. 
Row \textsf{(a)}
in \autoref{f:hst} shows that most (90\%) of the TIGRESS snapshots have
sufficient ionizing photon production rate to ionize  the  Reynolds layer. In fact, a
large fraction of snapshots (56\%) have $\Phi_{\rm i}$ that is high enough to
fully ionize even the {\it smooth} Dickey-Lockman layer,
$\Phi_{\rm i,min,D\mbox{-}L} = 3.4 \times 10^{50}\second^{-1}\kpc^{-2}$. 

Since the real gas distribution is clumpy and dense gas is highly correlated
with ionizing sources, however, the majority of ionizing photons are absorbed by
gas and dust near ionizing sources and the number of ionizing photons escaping
into the diffuse ISM is greatly reduced. For example, we find that the mean
value of ionizing photon flux passing through the planes $z = \pm 200\pc$
($\Phi_{\rm i,|z|=200\pc}$) is only $2.3 \times 10^{49}\second^{-1}\kpc^{-2}$, which is only 
6\% of time-averaged median $\Phi_{\rm i}$. Only 22\% of snapshots satisfy
$\Phi_{\rm i,|z|=200\pc} > 3.2\times 10^{49}\second^{-1}\kpc^{-2}$, which would be the
minimum required to maintain the ionization profile described by 
\autoref{eq:Reynolds} at
$|z| > 200 \pc$. 

\rev{ 
\HII\ region dynamics (not included in the current simulations) are likely to aid the 
escape of radiation from star-forming clouds, so that a larger proportion of 
$\Phi_{\rm i}$ would emerge from the midplane than 
we have found. Recent numerical simulations of
individual molecular clouds have shown that radiation feedback from massive stars plays a key role in driving gas dispersal on the scale of tens of parsecs \citep[e.g.,][]{walch2012,dale2012,dale2013,kimjg2018,haid2019,he2019,kimm2019,gonzalez2020}.
In particular, \citet{kim2019} showed that a significant fraction of ionizing
  photons escape on a short timescale ($\lesssim 3\Myr$) through low-density
  channels created by stellar feedback and turbulence, which can boost the
  photon budget to ionize DIG.}


In addition to a lack of photons, and perhaps more importantly, we believe that the
discrepancy between our median WIM profile and the observational estimate
is caused by the fact that there is simply not enough material to be ionized at
large $|z|$: the mean profile of $\langle \nH \rangle$ from our simulation is
lower than the observational constraints by a factor $3$--$5$ at $1$--$2\kpc$.
One possible reason for this discrepancy is that the TIGRESS simulation
underestimates outflows, potentially because effects of cosmic rays have not
been included. A second possibility is that inflowing extragalactic gas, not
captured in TIGRESS, is responsible for most of the extended WIM.  In \autoref{s:Haprofile},  we previously noted that the deficit of blueshifted  \Halpha\  in our synthetic profile (compared to WHAM) could potentially be due to missing extragalactic inflow.

It is also possible that the present state of the DIG in the local Milky Way
is atypical. Indeed, although the ionized gas content is insufficient to match
the observations for the majority of snapshots, we find that a small fraction of
snapshots have vertical profiles that are in good agreement with observational
constraints. For example, snapshots at $t \sim 550$--$570 \Myr$ have substantial
amounts of warm fountain gas at high altitudes lifted up by SN feedback from
previous generation of star formation. The ongoing star formation produces
sufficient photons, and channels are available for their escape, such that an
extended DIG layer comparable to the observed Reynolds layer is present.
Profiles from a snapshot at time $t=568\Myr$ shown in the bottom panel of
\autoref{f:zprofobs} is an example of such case. (see also
\autoref{f:snapshot2}).

Lastly, it has long been suggested that runaway OB stars may act as effective ionization
source of DIG \citep[e.g.][]{heiles1987,rand1993}. Runaways stars are flung at
high speeds from their birthplace by dynamical encounters in dense
stellar systems \citep[e.g.,][]{poveda67,fujii11} or by explosion of a companion
star in a binary system \citep[e.g.,][]{blaauw61,portegieszwart00}). Once they
move to high-altitude, low-density regions, ionizing photons emitted by these runaways
would more easily escape the galaxy and ionize warm neutral gas along the way
\citep[e.g.,][]{conroy2012}.

We find that runaways that represent binary 
companions have a negligible impact on
the ionization state, as shown in 
\autoref{appendix}.
However, we caution that the ionizing
photon rate that we used is likely an underestimate because all of runaways
modeled in our simulation are secondaries in binary systems, which are mostly
B-type (or late O-type) stars. Dynamically ejected runaways are likely to be
younger and more massive than binary runaways and produce more ionizing photons.
We also note that our approach to calculating the rate of ionizing photons from
runaways is not internally consistent because the SN rate as well as the
mass-luminosity relation are taken from stellar evolution and population
synthesis models containing only single stars \citep{leitherer1999,bruzual2003}.
If the effects of binary interaction is included, the ionizing photon rate at
late stage of stellar evolution could be boosted by several orders of magnitude
\citep[e.g.,][]{gotberg2019}.

\subsection{Comparison with Observations: the WIM scale height}

We measure a time-averaged mean scale height of the warm ionized gas 
$H_{\rm w,i} = 556\pm186 \pc$. This mean value is in good agreement with some
observed measurements of the Milky Way WIM, \rev{allowing for typical uncertainties of 
$\sim 100$--$200$ pc (\citealt{nordgren1992} measure a scale 
height of 670 pc, while \citealt{peterson2002} find a scale height of 830 pc). However, many empirical 
estimates of the average WIM scale  height are larger: 
\citealt{taylor1993}, \citealt{savage1990},  \citealt{reynolds1991a}, and \citealt{berkhuijsen2008} give a scale height of $\sim 900$ pc, while \citealt{gomez2001} measures a scale height of $\sim 1100$ pc, and 
\citealt{gaensler2008} measures a scale height of $\sim 1800$ pc.
The possible reasons given above for our  discrepancy with the overall ``Reynolds'' profile could potentially also explain why our measured WIM scale height is  smaller than most empirical estimates.
}


Under the simplistic assumption that a sample of Milky Way-like galaxies
should be similar to the ensemble obtained via evolution of the    TIGRESS box, our
results on the distribution of \Halpha\ scale height (as presented in
\autoref{f:hnesq}) can be compared to scale height measurements in nearby disk
galaxies. The scale height of the DIG in an external galaxy is measured
significantly differently from internal scale height measurements of the Milky
Way, both due to constraints on the nature of data collected (via \Halpha\ in
integral field unit spectroscopy or narrowband imaging) and due to technical
constraints (e.g., the influence of the point spread function on the observed
scale height). To account for this, we also compute scale heights
wherein an exponential profile is fit to the $\nesq$ vertical profile at
$|z|>1 \kpc$ (see \autoref{s:scaleheight}).

We find that our
distribution of \Halpha{} scale heights is generally 
similar to observed distributions, but because we have not attempted
to simulate the effects of the point spread function (PSF, which
often have extended low surface brightness wings), because our simulation 
aims to reproduce only Solar Neighborhood conditions and because 
samples of \Halpha{} scale heights remain relatively small, we 
cannot make a strict statement of (in)consistency from this
comparison.

We find a median exponential-fit $\nesq$ scale height of
$0.44^{+0.33}_{-0.15}\kpc$, here reporting the inner 68\% of the distribution in
order to compare to literature observations.
The distribution of measured \Halpha\ scale heights for a sample of edge-on disk
galaxies presented in \cite{levy2019} find a median scale height of
$0.8^{+0.7}_{-0.4}\kpc$, with the maximum likelihood scale height at
$\sim 0.5 \kpc$ and an extended tail towards scale heights of larger than
$1 \kpc$, much like our distribution of exponential-fit scale heights in
\autoref{f:hnesq}. \cite{bizyaev2017} find a somewhat
higher median \Halpha{} scale height of $1.2\pm 0.5 \kpc$ for a sample of
67 edge-on galaxies observed by MaNGA. 
Finally, \citet{jo2018} observes a mean \Halpha{} scale
of $430\pm50 \pc$ for a sample of edge-on galaxies at
$d_L\lesssim25$ Mpc, comparable to the mean of
the distribution of our exponential-fit scale heights ($535.4\pm10. \pc$).

As previously shown in \autoref{s:scaleheight}, we find no strong temporal correlation
between the global SFR density of the simulation and the exponential-fit $\nesq$
scale height. In the literature, a positive correlation has been reported
between \Halpha\ luminosity and \Halpha\ scale height \citep{bizyaev2017}, while
a weak anticorrelation has been reported between the extraplanar midplane EM
\footnote{Extrapolated to $z=0$, the reported values are those associated with the outer
exponential profile in a two-exponential fit.} and the \Halpha\ scale height
\citep{miller2003}. We again however emphasize that our simulations are
initialized with Solar neighborhood-like conditions, and cannot be interpreted
as direct analogs to external galaxy samples which may have a wider range of
conditions than presented by the time-varying state in our single TIGRESS
simulation.

\subsection{The Clumping Correction Factor}\label{s:clumping}

\begin{figure*}
\center{\includegraphics[width=\linewidth]{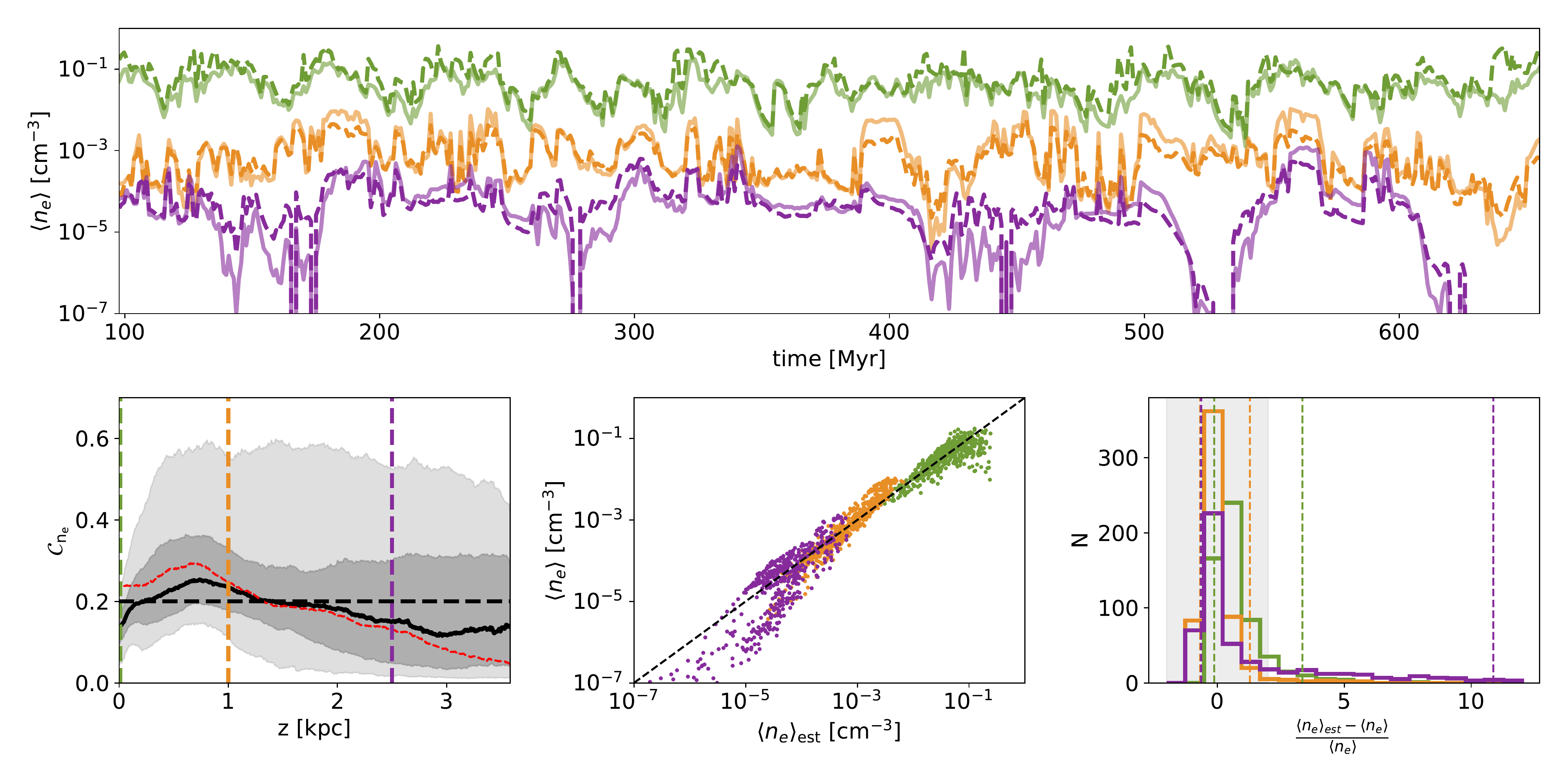}}
\caption{\textit{Top:} the time evolution of the average $\nelec$ at the
  midplane (green), $1\kpc$ above the midplane (orange), and $2.5\kpc$ above the
  midplane (purple). For each, the solid curves show the true values of
  $\langle \nelec \rangle$ and the dashed curves show the estimated average
  $\langle \nelec \rangle_{\rm est}$ using the observable $\langle n_e^2 \rangle$ and the 
  height-averaged, time-averaged median
  clumping correction factor, $C_{\nelec}^{50}=0.2$. \textit{Bottom Left:} 
  the clumping correction factor as a function of height, $C_{\nelec}(z)$
  (as given by \autoref{eq:nest}). The solid black curve shows the time-averaged median
  value of the clumping correction factor, while the dashed horizontal line shows the
  time-averaged, height-averaged median value of the clumping correction factor. The grey
  shaded regions show the 25$^{\rm th}$ to 75$^{\rm th}$ and 5$^{\rm th}$ to
  95$^{\rm th}$ percentile regions. The colored dashed lines show the heights
  that correspond to the time evolution plot at top. We also
  show the median profile of the square root of the warm ionized volume filling factor, $f_{V,\rm WIM}^{1/2}$, with a red dashed curve. In the idealized case 
  of constant $\nelec$ within  fully-ionized portions of the volume, 
  $C_{\nelec{}} = f_{V,\rm WIM}^{1/2}$.
  \textit{Bottom Middle:} the
  relation between estimated electron density
  ($\langle \nelec \rangle_{\rm est}$) and the true $\langle \nelec \rangle$,
  colored by height. \textit{Bottom Right:} the distribution of fractional error
  incurred by estimating $\langle \nelec \rangle$ using the time-averaged, height-averaged median
  value of the clumping correction factor. The grey shaded region is bounded by [-2,2]. The
  histograms show the fractional error distribution at the midplane (green),
  $1 \kpc$ (orange), and $2.5 \kpc$ (purple). The dashed vertical lines show the
  5$^{\rm th}$ and 95$^{\rm th}$ percentiles of the distribution for each of the
  heights above.}
\label{clumpinesscalibration}\vspace{30pt}
\end{figure*}

\Halpha\ observations of edge-on disk galaxies have been used to infer the
physical properties of the DIG such as electron density and total mass/column density
\citep[e.g.,][]{dettmar1990,rossa2000,boettcher2019}. Because the \Halpha\
surface brightness (or EM) is proportional to the integral of squared electron
density along the line of sight, however, deriving these quantities from the EM
requires a knowledge of the spatial distribution of ionized gas as well as the
effective path length through the galaxy.
Simulations allow us to directly measure the ``clumping correction factor'' needed to
convert EM into electron density. In addition to providing a calibration,
simulations also allow us to gauge the uncertainties incurred by assuming a
constant value of the clumping correction factor in a dynamic and varying system.

Here, we compute the clumping correction factor of DIG as a function of $z$, and use this
as a multiplicative factor to determine the value of the (line-of-sight averaged)
electron density from an observed EM. We
also provide a simple analytic expression for the effective path length in terms
of disk scale length and local radius. These results are intended to aid in
obtaining estimates and expected errors for the content
of the DIG based on observations of external galaxies.

We define the clumping correction factor at height $z$ as
\begin{align}
  \mathcal{C}_{\nelec} & \equiv \sqrt{\dfrac{(\int \nelec\Thetaw dA/A)^2}{\int \nelec^2 \Thetaw dA/A}} \\
                        & = \sqrt{\frac{\mean{\nelec}{}^2}{\mean{\nelec^2}{}}}\,, \label{eq:C}
\end{align}
where the brackets $\langle \rangle$ denote the $x$-$y$ area average and $\Thetaw{}$ is
a top hat function that filters warm gas with
$5\times 10^3 \Kel < T < 2\times 10^4 \Kel$. In the idealized case where all WIM
clouds are fully ionized and have the same electron density, the clumping correction factor and volume filling factor at a given height are related:
$\mathcal{C}_{\nelec}=f_{V,{\rm WIM}}^{1/2}$, where
$f_{V,{\rm WIM}}=\int x_{\rm i}\Thetaw dA/A$.

From our simulations, we compute statistics of the clumping correction factor
${\mathcal{C}}_{\nelec}$ as a function of height off the midplane ($z$) and
time. ${\mathcal{C}}_{\nelec}$ may then be used with a known value of
$\langle \nesq \rangle$ to predict the area-averaged mean electron density at
$z$. 
Let us define the instantaneous average of EM over the local radial direction ($x$) in
the plane of the sky at height $z$ as $\mean{{\rm EM(z)}}{x} \equiv \int \nesq dx dy/L_x = \langle \nesq \rangle L_y$.
Then, from \autoref{eq:C} we have
\begin{equation} \label{eq:nest}
  \langle \nelec \rangle_{\rm est}(z) = \mathcal{C}_{\nelec}(z)
  \sqrt{\dfrac{\mean{\rm EM(z)}{x}}{L_y}}\,. 
\end{equation}

The bottom-left panel of \autoref{clumpinesscalibration} shows the time-averaged
median of $\mathcal{C}_{\nelec}(z)$ from our simulation, as well
as the 25$^{\rm th}$ to 75$^{\rm th}$ and 5$^{\rm th}$ to 95$^{\rm th}$
percentiles. Evidently, $\mathcal{C}_{\nelec}(z)$ is largely
constant as a function of $z$, with an
overall (time-averaged, height-averaged) median value of $C_{\nelec{}}^{50} = 0.2$ (and a 25\thh{} and 75\thh{}
percentile value of $C_{\nelec}^{25}=0.11$ and $C_{\nelec}^{75}=0.31$). We also find
that the clumping correction factor does not vary significantly with electron density or
total gas density. 
We show the application of this clumping correction factor as a function of time for gas at
the midplane (green), $1\kpc$ (orange), and $2.5 \kpc$ (purple) in the top panel of
\autoref{clumpinesscalibration}. Here, the solid curves show the true
value of $\langle \nelec\rangle$ and the dashed curves show the estimated
value $\langle n_{\rm e}\rangle_{\rm est}$ using the constant 
calibration factor of $\mathcal{C}_{\nelec}^{50}=0.2$. 

Using this median value in \autoref{eq:nest}, the
bottom-middle panel of \autoref{clumpinesscalibration} compares the actual
$\langle \nelec \rangle$ with the estimated value for all snapshots at
$z=0,1,2.5 \kpc$, colored by height. The distribution of fractional error is then
shown in the bottom-right panel of \autoref{clumpinesscalibration}. This shows
that using a time-averaged, height-averaged median value of $\mathcal{C}_{\nelec}=0.2$ we
can successfully convert from $\langle \nelec^2 \rangle$ to
$\langle \nelec \rangle$ within a factor of 2 for 95\% of snapshots at
$|z| \sim 1\kpc$.

For application to observed galaxies, we suppose that an observer wants to make
an estimate of the local electron density of DIG along the line of sight in an
edge-on galaxy, such that the plane of the sky is the $x$-$z$ plane ($x$ is the
local radial direction). To apply \autoref{eq:nest}, the known $L_y$ in the simulation must be replaced with
${\cal \ell}_y$, an effective path length along the line of sight.

To illustrate the idea, let us assume that the electron density
$\tilde{n}_{\rm e}$, averaged over a small volume (but sufficiently
representative of the local condition),
is an exponential function of the galactocentric radius
$\tilde{n}_{\rm e}(R,z) = \tilde{n}_{\rm e,0}(z) e^{-R/R_d}$, where $R_d$ is the
radial scale length\footnote{We note that in their estimate of electron density,
  \citet{boettcher2019} do not include radial density dependence based on the
  fact that DIG in some galaxies exhibit highly filamentary morphology
  \citep{collins2000,heald2006}.}. If the clumping correction factor $\mathcal{C}_{\nelec}$
does not vary with $R$, the EM at some (projected) galactocentric radius $R$ and
at some height $z$ can be written as
\begin{align}
    {\rm EM}(R,z) & = \int_{-\infty}^{\infty} \mathcal{C}_{n_{\rm e}}^{-2}
                    \tilde{n}_{\rm e,0}^2(z)  e^{-\frac{2}{R_d}\sqrt{R^2 + y^2}}
                    dy \\ 
    & \equiv \mathcal{C}_{n_{\rm e}}^{-2} \tilde{n}_{\rm e}^2(R,z) \ell_y
\end{align}
where the effective path length
$\ell_y = 2R \int_0^{\infty} e^{-q \sqrt{1 + s^2}} ds$ with $q = 2R/R_d$. The
integral can be approximated as
\begin{equation}
\int_0^{\infty} e^{-q \sqrt{1 + s^2}} ds \approx \frac{e^{-q}(1+q)}{q(1 + 0.2q)}\,,
\end{equation}
which is accurate to within $\sim 10\%$ for $q < 10$.\footnote{We arrive at this
  approximation by integrating in the regimes of small and large $s$, i.e.
  $\int_0^\infty e^{-q\sqrt{1+s^2}}ds \approx \int_0^1 e^{-q}ds + \int_1^\infty
  e^{-qs}ds = e^{-q}q/(q+1)$ and introducing an additional fudge factor
  $(1 + 0.2q)^{-1}$.} The estimate of electron density is then
\begin{equation}\label{e:nest_obs}
  \tilde{n}_{\rm e}(R,z) \approx \mathcal{C}_{\nelec} \left(\frac{{\rm
        EM}(R,z)}{R_d}\right)^{\frac{1}{2}}\left(\frac{1 + 0.4\frac{R}{R_d}}{1 +
      \frac{2R}{R_d}}\right)^{\frac{1}{2}}e^{R/R_d} 
\end{equation}
for the adopted value of $\mathcal{C}_{\nelec{}}$. In practice, $R_d$ can be estimated
from the \Halpha\ scale length of the disk, i.e., $R_d = 2 R_{{\rm H}\alpha}$.
While the measured EM (or \Halpha\ surface brightness) would be a weighted
average of ${\rm EM}(R,z)$ over a finite area due to PSF effects, for spatially
resolved edge-on galaxies \autoref{e:nest_obs} would provide a reasonably good estimate
of the density of DIG.

\citet{berkhuijsen2008} estimated a volume filling factor of the WIM  $f_{V,{\rm WIM}}\sim 0.1$, based on EM and DM to pulsars with distances known within $50\%$ and assuming uniform $n_e$. The implied clumping correction factor would then be $\mathcal{C}_{\nelec}=0.3$, slightly larger than the value obtained from our simulation.  Finally, we note that although $\mathcal{C}_{\nelec}$ is identically equal to $(f_{V,{\rm WIM}})^{1/2}$ only in the case of constant $\nelec$ in fully-ionized regions, we find that these quantities are in fact quite close to each other.  In \autoref{f:mzprofiles}(b), we presented the median WIM volume filling factor; in the lower-left panel of \autoref{clumpinesscalibration} we now overplot the square root of this in red.

\section{Summary}\label{s:conc}

In this study, we have applied radiation-transfer post-processing to TIGRESS MHD
simulation outputs in order to study the properties of the WIM.
The sources of ionizing (and FUV) photons for the radiative transfer are star
particles that have formed as a result of self-gravitating collapse. To our
knowledge, this is the first such radiative transfer model in which
star formation, SN rates and locations, the (global) FUV heating rate, and the
properties of ionizing sources are self-consistently determined. With a
self-consistent spatio-temporal correlation of SNe and ISM gas, the overall
vertical distribution of warm gas and the network of porous (hot ISM) channels
that allow photons to propagate over kpc scales are more realistic than in
previous work, and the photon input rate is
set by the time-varying star formation rate rather than being a free
parameter.

The TIGRESS model that we consider represents a galactic environment similar to
the local Milky Way. The median ionizing photon production
rate is $\Phi_{\rm i} = 4.0 \times 10^{50} \second^{-1} \kpc^{-2}$, although this
varies significantly as only $\sim 3$ sources account for 90\% of the ionizing
photons at any one time (see \autoref{f:numsrc}). The mean escape fraction of ionizing photons from the
galaxy is only $1.1\%$;
A much larger fraction ($\sim 22\%$) of non-ionizing
photons escape from the galaxy 
(\autoref{f:hst}c).

Most of the ionizing photons are absorbed by dense gas
($\nH > 1 \cm^{-3}$) near the midplane. 
Overall, approximately half of ionizing photons are
absorbed by neutral hydrogen and half by dust 
(\autoref{f:hst}d); absorption by
neutral hydrogen is completely
dominant at high $|z|$ where the ionization parameter is low
($U=10^{-5}-10^{-3}$, see \autoref{dustabs}). 

Nevertheless, diffuse warm ionized gas is abundant
far from the midplane, with $40\%$ of the ionized gas mass found at
$|z| > 200\pc$, and a peak in the WIM volume filling factor at
$|z| \sim 0.5 - 0.8 \kpc$ (\autoref{f:mzprofiles} \textsf{(c)} and \textsf{(b)}, 
respectively). The proportion of warm gas that is ionized increases with $|z|$ (\autoref{f:mzprofiles}\textsf{(f)}). There is a positive correlation between volume filling factor of ionized gas and SFR (\autoref{fvsfr}). 

To make a quantitative comparison to observations, we fit the high-$|z|$ ($>1\kpc$) EM to an exponential. Our simulation shows 
large temporal variability in this exponential-fit scale height, with
a typical value of $h(\nesq) \sim 400\pc$ and a tail that extends up
to $\sim 2$\kpc{} (\autoref{f:hnesq}). 
This is comparable to estimates of the DIG scale height in both the Milky
Way and external edge-on galaxies, though small observational sample
sizes and the applicability of Solar Neighborhood conditions to those
samples make a direct comparison uncertain. 

Our simulations also allow us to assess the
``clumping'' of ionized gas. At $|z|>200\pc$, most of the \Halpha\ emission is
from ionized gas clouds with local density $0.1$-$1 \cm^{-3}$ (\autoref{f:nedist}), although the
horizontal average $\langle \nelec\rangle$ and the ``typical'' $n_e = \langle \nelec\rangle/f_{V,{\rm WIM}}$ are much lower (panels
\textsf{(c)} and \textsf{(d)}, respectively, of \autoref{f:mzprofiles}). 

We find that the mean
(volume-averaged) electron density in the DIG is related to the mean value of the square of
the electron density by
$\langle \nelec \rangle = \mathcal{C}_{\nelec} \sqrt{\langle \nesq \rangle}$ for
$\mathcal{C}_{\nelec} \sim 0.2$ (see \autoref{s:clumping}). 
In our simulations, this clumping calibration factor is
remarkably independent of time and of height above the midplane. The numerical
calibration of clumping should be useful in converting EM from observations of
\Halpha\ surface brightness (or other WIM tracers) in edge-on galaxies to mean
density of the DIG.
For the case of disk with density declining exponentially with radius, we give an approximate prescription for obtaining $\nelec{}(R,z)$ from $EM(R,z)$,  the exponential scale length $R_d$, and the 
distance $R$ from the center of the galaxy 
(\autoref{e:nest_obs}).


When observed ``face-on'' (along the z-axis), the EM distribution of the WIM in the disk is
centered on $\sim 0.1 \pc \cm^{-3}$ (see \autoref{f:emhst}), 
though the breadth of the distribution
increases for smaller effective beam size (\autoref{f:EMbeam}). 
Overall, even though nearly half of
ionized gas in our simulations comes from high-altitude regions
($|z| > 200 \pc$), only $\sim 5\%$ of the \Halpha\ emission is produced there, due to the low gas density.

In comparing our synthetic line profiles to WHAM observations, we find that the
shape and normalization of the time-averaged redshifted (outflowing) side
matches well, but that the simulation does not produce sufficient
\Halpha{} emission at blueshifted velocities (inflowing, see 
\autoref{velprofile}).
We also find that the mean electron density inferred from DM of pulsars at high
$|z|\sim \kpc{}$ significantly exceeds the level in our simulations (\autoref{f:zprofobs}). Taken
together, this raises the intriguing possibility that much of the Reynolds layer is an
ionized extragalactic inflow. An alternative possibility, of course, is that the
TIGRESS simulations have significantly underestimated the mass involved in
fountain flows, and at the current time the fountain is mostly in an inflow
episode. However, our mean profiles of neutral gas match the Dickey-Lockman 
profile
inferred from 21 cm out to $\sim 3\kpc$ (\autoref{f:zprofobs}). We also note that
for some individual epochs in our simulation, the instantaneous $n_e(z)$ profile
matches that inferred from pulsar DM very well 
(as discussed in \autoref{s:obs}).

The high-velocity portions of our synthetic \Halpha\ profiles are exponential (\autoref{velprofile}), consistent with the underlying exponential distribution of mass with velocity in  extraplanar ``fountain'' gas.  
At most times in our models, 
more than half of the high-velocity (defined as $|v|>50\kms$) ionized gas is 
located at high altitudes (\autoref{highvelcontrib}). In cases where the \Halpha\ wing
emission at $v>50\kms$ exceeds $\sim$2\% of the total \Halpha{} emission, the majority of the wing emission originates from high-altitude gas.
Periods of stronger wing emission correspond to strong outflows, and are
time-delayed relative to strong bursts of star formation.

Finally, we remark that while the present study offers a significant advance
over previous work, it is not without limitations. In particular, the ionization
state is computed via post-processing, where for each snapshot, we evolve until
ionization equilibrium is reached. In reality, for low-density, high-altitude
regions, the radiation field (or source visibility) may change more rapidly than
the recombination rate, so that the true ionization level may be higher than we
estimate. For simplicity, we have adopted a constant temperature for the ionized
gas, but to make more comparisons with multi-line observations heating and
cooling of ionized gas must be included self-consistently. Inclusion of cosmic
rays that dynamically interact with the gas could potentially enhance the mass
of gas in the extraplanar region. \rev{As mentioned in
  \autoref{s:discrepancies}, resolving cloud-scale feedback processes before the
  first SNe could increase the fraction of ionizing photons escaping
  into the diffuse ISM.} In the future, it will be valuable to include
time-dependent radiative transfer 
\rev{ and adaptive mesh refinement} as part of the MHD simulation.

The present study deals only with Solar
neighborhood-like conditions. It is also 
of interest for the future to extend the
same kind of analysis to other galactic environments; simulations with
conditions representative of galactic centers and disk outskirts will allow for
fuller predictions of the state of DIG in disk galaxies.

\acknowledgements

\rev{
We thank the anonymous referee for a helpful review of this work.}
J.-G.K. acknowledges support from the Lyman Spitzer, Jr. Postdoctoral Fellowship
at Princeton University.
This project was partly supported by NASA under ATP grant NNX17AG26G to E.C.O.
The work of C.-G.K. was partly supported by a grant from the Simons Foundation
(CCA 528307, E.C.O.). Resources supporting this work were provided in part by
the NASA High-End Computing (HEC) Program through the NASA Advanced
Supercomputing (NAS) Division at Ames Research Center and in part by the
Princeton Institute for Computational Science and Engineering (PICSciE) and the
Office of Information Technology's High Performance Computing Center.

\software{Astropy \citep{astropy:2013, astropy:2018}, matplotlib \citep{Hunter:2007}, SciPy \citep{jones_scipy_2001}, the IPython package \citep{PER-GRA:2007}, NumPy \citep{van2011numpy}, and Scikit-learn \citep{mckinney}}

\bibliography{refs.bib}

\appendix

\section{Effect of Runaways}\label{appendix}

Binary runaway OB stars ejected from stellar clusters after a supernova can, in
principle, act as sources of ionizing radiation at large distances from the disk
(see \autoref{f:snapshot1}, leftmost panel). Because TIGRESS tracks such
runaways, we can examine whether such stars make a significant contribution to
the DIG. The main sequence lifetime of each runaway is chosen to be consistent
with the specific SNe rate (bottom panel, \autoref{f:sb99}) -- a full
description of the prescription for this assignment is given in \cite{kim2017}.
The ionizing photon luminosity of a runaway is then computed from its main
sequence lifetime using the main sequence lifetime-mass and mass-luminosity
relations given in \cite{parravano2003}. Since the EUV luminosity drops to
relatively negligible levels at $t>10 \Myr$, we consider a maximum of 20 brightest runaways
as sources of ionizing radiation simultaneously. For most of snapshots, this accounts for more than 90\% of the
total ionizing photon rate of all runaways.

\begin{figure*}
\center{\includegraphics[width=\linewidth]{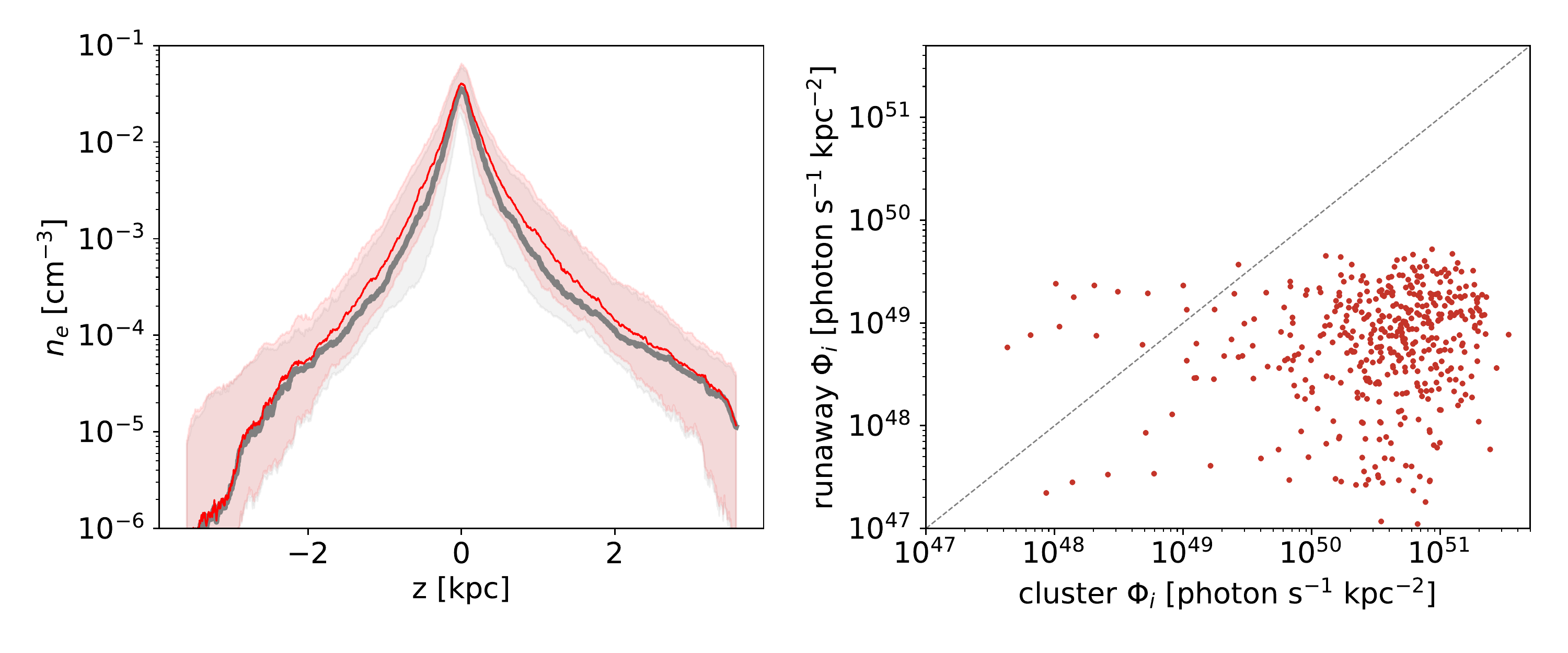}}
\caption{\textit{Left:} Time-averaged $\nelec{}$ vertical profiles when the
  contribution of runaways is included (red) and when runaways are neglected
  (grey). As in \autoref{f:mzprofiles}, the solid curves show the median
  profile, while the shaded regions show the 25\thh{} and 75\thh{} percentiles.
  \textit{Right:} the specific ionizing photon rate $\Phi_{\rm i}$ of clusters
  versus $\Phi_{\rm i}$ from runaways. For most snapshots, runaways contribute
  about 1\% as many ionizing photons as clusters.}
\label{f:runawaydiag} \vspace{30pt}
\end{figure*}

In \autoref{f:runawaydiag}, we show diagnostics of the effect of runaways on the ionizing photon rate in the simulation box and structure
 of the WIM. In the left panel, the grey curve
show the ionizing photon rate per unit area when runaways are not considered as
ionizing sources, while the red curve show the results when runaways are
included as sources of ionizing radiation. The right panel shows the ionizing
photon rate for clusters versus that from runaways for each snapshot. The median
ionizing photon rate from runaways is 1.7\% that of clusters -- though the rate
of ionizing photons from runaways is as high as that from clusters in 3\% of
snapshots, the overall contribution of runaways to the ionizing photon rate is
smaller than that of clusters by several orders of magnitude.
 
The effect of runaways is small relative to the variation in time, and does not
induce a significant systematic change in the structure of the DIG. We find that
the inclusion of runaways induces a change in the area-averaged WIM electron
density that is small compared to the fluctuations we observe due to time
evolution, as shown in the left panel of \autoref{f:runawaydiag}. Thus, we
choose to exclude runaways as sources of ionizing radiation for this work.

\end{document}